\documentclass[12pt]{iopart}
\usepackage{graphicx}
\usepackage{mathrsfs}  
\usepackage{amsfonts}
\usepackage{eqnarray}
\usepackage[colorlinks=true,
            linkcolor=blue,
            urlcolor=blue,
            citecolor=blue]{hyperref} 
\usepackage[caption=false]{subfig}
\usepackage{enumitem}
\usepackage{cleveref}
\crefformat{footnote}{#2\footnotemark[#1]#3}
\newcommand{\hdots}{\dots}
\newcommand{\be}{\begin{equation}}
\newcommand{\ee}{\end{equation}}
\newcommand{\eqref}[1]{\eref{#1}}
\newcommand{\ba}{\begin{array}}
\newcommand{\ea}{\end{array}}
\newcommand{\prj}[1]{\ensuremath{| #1 \rangle \langle #1 |}}
\newcommand{\bra}[1]{\ensuremath{( #1 |}}
\newcommand{\ket}[1]{\ensuremath{| #1 )}}
\renewcommand{\bullet}{\,}
\usepackage{bm}

\begin{document}

\title{Exploring complete positivity in hierarchy equations of motion}
\author{B Witt$^{1,2}$, \L\ Rudnicki$^{1}$, Y Tanimura$^{2,3}$, F Mintert$^{1,2}$}

\address{$^1$ Department of Physics, Imperial College London, London SW7 2AZ, United Kingdom}
\address{$^2$ Freiburg Institute for Advanced Studies, Albert-Ludwigs-Universit\"at, Albertstra\ss e 19, 79104 Freiburg, Germany}
\address{$^3$ Department of Chemistry, Kyoto University, Kyoto, Japan}
\ead{b.witt14@ic.ac.uk}
\vspace{10pt}

\begin{abstract}
We derive a purely algebraic framework for the identification of hierarchy equations of motion that induce completely positive dynamics and demonstrate the applicability of our approach with several examples.
We find bounds on the violation of complete positivity for microscopically derived hierarchy equations of motion and construct well-behaved phenomenological models with strongly non-Markovian revivals of quantum coherence.
\end{abstract}
\newpage

\section{Introduction}
Every quantum system inevitably interacts with its environment.
This interaction contributes to the emergence of classicality, represents the major challenge in the development of quantum information technological hardware \cite{Devoret2013}, but can also be exploited for advanced means to control quantum systems like laser cooling \cite{Phillips1992} or dissipative state preparation \cite{Kraus2008}.

The central difficulty in describing open quantum systems resides in the mere size of an environment.
Most approaches aim at the reduction of such environments to their aspects that are most relevant for the system dynamics.
This ranges from effective descriptions in terms of the system's degrees of freedom only \cite{Lindblad1976, Gorini1976}, via approaches that explicitly include the most relevant environmental degrees of freedom \cite{Cederbaum2005, Iles-smith2014, Chin2010a} to numerically expensive treatments that aim at a potentially exact description \cite{Tanimura1989a,Tanimura2014,Tanimura1990,Tanimura2015,Tanimura2006}.
In practice, one needs to take a compromise between accuracy and numerical effort;
a comparatively simple effective description can help to develop a more intuitive model, but may also fail to identify more subtle features.

A completely different approach is based on a phenomenological modelling of open quantum dynamics that promises a better understanding and deeper insights into the underlying physics.
The hallmark is provided by the Kossakowski-Lindblad equation \cite{Lindblad1976, Gorini1976} that allows us to construct models that respect the probabilistic interpretation of quantum mechanics:
Markovian Lindblad master equations ensure the property of {\it complete positivity} such that all quantities that are interpreted as probabilities are always non-negative. 
Furthermore, it permits an interpretation of the environmental effects in terms of rates and elementary processes. The intuitive comprehension that comes along with this model suffers the disadvantage of a strongly limited applicability. Many systems that are actively investigated, like quantum dots \cite{Fischbein2005}, light-harvesting systems \cite{Collini2010}, or photonic band-gap materials \cite{DeVega2005}, are characterized by a rather strong system-environment coupling giving rise to non-Markovian effects like the back-flow of information that are not captured by this model.

For that reason, it is desirable to find a framework for the description of non-Markovian processes that ensures complete positivity and is in the best case even phenomenologically accessible.
Complete positivity is not only crucial for physically sensible dynamics, but understanding the very conditions under which the equation of motion induces completely positive (CP) processes must also be considered a crucial step towards a phenomenological understanding.
However, the addition of memory to the system dynamics gives rise to substantial complications \cite{Barnett2001} in particular with respect to complete positivity.
Most approaches to this problem focus on integral equations over a memory kernel and for certain classes like semi-Markov processes \cite{Breuer2009b,Chruscinski2012}, collision-model-based approaches \cite{Ciccarello2013} or continuous time quantum random walks \cite{Budini2004} conditions for complete positivity have been found.
Some of these classes can also be covered by generalized sufficient conditions that have been obtained recently \cite{Chruscinski2016a}.
Eventually, the post-Markovian master equation \cite{Shabani2005} is capable to describe CP non-Markovian dynamics by interpolating between the generalized measurement interpretation of the Kraus operator sum and the notion of a continuous measurement for Markovian processes but eventually also imposes conditions on the memory kernel.

In this article we propose a different take on the question of CP non-Markovian dynamics.
Rather than by integral forms over a memory kernel, our approach is inspired by the framework of hierarchical equations of motion (HEOM) \cite{Tanimura1989a} that employ -- in addition to the system density matrix $\varrho_1(t)$ -- a set of auxiliary operators $\varrho_i(t)$ ($i=2,\hdots, n$) that contain information about the environment and system-environment correlation.
Equations of motion of the form
\be
\dot\varrho_{i}(t)=\sum_{j=0}^n{\mathcal L}_{ij}\,\varrho_j(t)\ ,\ (i=1,2,\hdots, n)
\label{eq:hierarchy}
\ee
have been derived based on microscopic models for a wide range of
systems including light-harvesting complexes \cite{Ishizaki2009,Strumpfer2009},
molecular transistors \cite{Santamore2013},
as well as electrons in tunneling junctions \cite{Sakurai2013} and organic heterojunctions exposed to a phonon bath \cite{Yao2014}. In particular they have proven to be very reliable in the regime where comparable intra-system and system-environment interactions prevent approximations based on separations of time-scales.

We will start out assuming the structure~\eqref{eq:hierarchy} with a finite-dimensional system state $\varrho_1(t)$ and target the identification of conditions on ${\mathcal L}_{ij}$ such that the {\it dynamical map} $\Lambda_1(t)$ with  $t \ge 0$ defined via $\varrho_1(t)=\Lambda_1(t)\,\varrho_1(0)$
is CP.
In the case of $n=1$, i.e.\ equations of motion for the system state only, it is well established that this condition is satisfied if and only if ${\mathcal L}_{11}$ can be written in Lindblad form \cite{Lindblad1976, Gorini1976}.
A generalization of this criterion to hierarchy equations with $n>1$ that also permits an incorporation of non-Markovian dynamics is, however, not  known.

In this article, we propose a new method by means of which sufficient conditions for CP dynamics can be derived for HEOMs.
In some cases, such conditions can be obtained analytically, whereas highly efficient numerical methods are in place whenever an analytical solution is not feasible.
In \sref{sec:the_algebraic_framework}, the formal and algebraic framework is set before a very intuitive geometrical interpretation of the procedure is given in \sref{sec:geometric_interpretation}.
In \sref{sec:simplification_of_condition_2_and_3}, we provide practical tools that ease an analytical examination as it is carried out for a couple of examples in \sref{sec:introductory_examples}.
More advanced techniques that aim for a numerical investigation of more complex problems are subject to \sref{sec:transformation_towards_simplified_geometries}, where we introduce a transformation that permits a universal applicability of the method (\sref{sec:transformation_of_the_extended_dynamical_map}) and demonstrate  how semi-definite programming can be employed for a highly efficient numerical examination of HEOMs (\sref{sec:semi-definite_programming}).
A truncation of a HEOM is often carried out in order to obtain approximate dynamics but can also give rise to equations that violate complete positivity.
That is why in \sref{sec:complete_positivity_after_an_initial_violation_time} and \ref{sec:lower_bound_for_the_eigenvalues_of_chi}, our approach is modified such that the binary classification into CP and non-CP dynamics is replaced by a quantification of the violation of complete positivity.
This permits the application of our method to an important example of a truncated HEOM that describes the spin-Boson model in \sref{sec:example_the_spin-boson_model}, before we conclude our results in \sref{sec:conclusion}.

\section{Deriving conditions for complete positivity}
\label{sec:monotone-based_conditions_for_complete_positivity}

Since complete positivity is a property of the dynamical map rather than of the system state, it is helpful to re-formulate \eqref{eq:hierarchy} in terms of time-dependent maps $\Lambda_i(t)$ that satisfy $\varrho_i(t)=\Lambda_i(t)\,\varrho_1(0)$.
The equations of motion for these dynamical maps are obtained from \eqref{eq:hierarchy} and read
\be
\label{eq:dynamical_map_equation_of_motion}
\dot\Lambda_{i}(t)=\sum_{j=1}^n{\mathcal L}_{ij}\,\Lambda_j(t)\ ,\ (i=1,2,\hdots, n)\ .
\ee
This set of equations can be expressed in a very concise form,
when we introduce the {\it extended dynamical map}
$\bm{\Lambda} = \sum_{i=1}^n \Lambda_{i}(t) \otimes \ket{i}$ with $\{\ket{i}\}$ being the orthogonal and normalised basis of an $n$-dimensional Hilbert space.
This notion of extended objects that are formed by operators is highly convenient and will in the course of this article be denoted by bold-faced symbols.
The scalar product on this extended space reads
\be
\langle \bm{\Lambda},\bm{\Gamma} \rangle \equiv \textrm{tr}(\bm{\Lambda}^\dagger\,\bm{\Gamma})= \sum_i \textrm{tr}(\Lambda^\dagger_i\,\Gamma_i)
\ee
for any two extended vectors $\bm{\Lambda}$ and $\bm{\Gamma}$.
In the same spirit, we also introduce the {\it extended state} $\bm{\varrho}(t) = \sum_{i=1}^n\varrho_i(t)\otimes\ket{i}$ and define the extended generator
$\bm{\mathcal{L}} = \sum_{i,j=1}^n \mathcal{L}_{ij} \otimes \ket{i}\bra{j}$, which permits to express \eqref{eq:dynamical_map_equation_of_motion} in the compact form
\be
\partial_t\bm{\Lambda}(t) = \bm{\mathcal{L}}\,\bm{\Lambda}(t)\ .\\
\label{eq:maphierarchy}
\ee
By means of this notation we are in the position to concisely describe the algebraic framework for the derivation of conditions for completely positive dynamics.
In practice, it is often convenient to work with Bloch-type representations, where the Bloch vector $\sum_ib_i\otimes\ket{i}$ of $\bm{\varrho}$ is defined in terms of the Bloch vectors $b_i$ of the operators $\varrho_i$,
{\it i.e.} vectors with components $[b_i]_j=\tr(\sigma_j\varrho_i)$ for $j=0,x,y,z$, where $\sigma_x$, $\sigma_y$ and $\sigma_z$ are the Pauli matrices and $\sigma_0 = \mathbb{I}$, or their generalisation to higher dimensional systems.
For extended maps $\bm{\Lambda}$ and generators $\bm{\mathcal{L}}$ the corresponding representations are defined analogously.

\subsection{The algebraic framework}
\label{sec:the_algebraic_framework}
Let us first detail the concept of valid maps. 
An extended dynamical map $\bm{\Lambda}(t)$ is valid if and only if the system map $\Lambda_1(t)$ is CP, i.e.\ it can be written as
\be
\Lambda_1(t)\, \varrho_1=\sum_{ij}\chi_{ij}(t)\ \mu_i\varrho_1\mu_j^\dagger\ ,
\label{eq:Kraus}
\ee
with a  positive semi-definite matrix $\chi(t)$ and a set of mutually orthonormal operators $\mu_i$.
Showing complete positivity of $\Lambda_1(t)$ is thus equivalent to proving positive semi-definiteness of $\chi(t)$.

The initial map $\Lambda_1(0)$ at the time $t=0$ is given by the identity map such that $\chi(0)$ has rank one.
The rank of a matrix provides information about the number of non-vanishing eigenvalues and does therefore also indicate the order of the highest non-vanishing elementary symmetric polynomial.
Those elementary symmetric polynomials read
\be
e_1(\chi) \equiv \textrm{tr}(\chi),
\hspace*{5px}
e_2(\chi) \equiv  [ \textrm{tr}^2(\chi) - \textrm{tr}(\chi^2) ] / 2,
\hspace*{5px}
\hdots,
\hspace*{5px}
e_r(\chi) \equiv \det(\chi)\ ,
\ee
where $r \times r$ is the dimension of $\chi$, and when $\chi$ has rank $k$, then $e_j(\chi)=0$ for all $j > k$.
For the initial matrix $\chi(0)$ one  obtains $e_1(\chi(0))=1$ and $e_j(\chi(0))=0$ for all $j>1$ but as time evolves, the eigenvalues of $\chi(t)$ will typically take non-vanishing values such that the rank of $\chi(t)$ increases.
We will now consider a time $t_p>0$ at which all non-trivial ({\it i.e.}\ not permanently vanishing) eigenvalues of $\chi(t)$ are strictly positive.
This is equivalent to $\chi(t_p)$ being positive semi-definite and having rank $h \equiv  \max(\textrm{rank}(\chi(t))\,|\,t \ge 0)$.
Any sign-flip of any eigenvalue of $\chi(t)$ for $t > t_p$ is then necessarily accompanied by a vanishing polynomial $e_h$, for why it is sufficient to ensure that $e_h(\chi(t))$ does never become  equal to zero for $t>t_p$ in order to show complete positivity from time $t_p$ on.
If no eigenvalue of $\chi(t)$ remains equal to zero ({\it i.e.}\ $h=r$), then $e_h$ is given by the determinant but even when one or more eigenvalues of $\chi(t)$ vanish permanently, the proposed method will still be applicable with $e_h$ being an elementary symmetric polynomial with lower degree.

In order to prove $e_h(\chi(t)) > 0$ for $t>t_p$, it is assumed (the assumption will be justified in \sref{sec:transformation_of_the_extended_dynamical_map}) that one can find a non-linear transformation of the coordinates $\bm{\Lambda}(t)$ such that $e_h(\chi(t))$ is quadratic in the extended map $\bm{\Lambda}(t)$ and can be written as
\be
\label{eq:representation_det_chi}
e_h(\chi(t))= \langle \bm{\Lambda}(0), {\bm S}\bullet \bm{\Lambda}(0) \rangle - \langle \bm{\Lambda}(t), {\bm S} \bullet \bm{\Lambda}(t)\rangle
\ee
with a suitable operator $\bm S$ and satisfy ${\bm S} = {\bm S}^\dagger$.

As we will show in the following
the dynamical map $\Lambda_1(t)$ is CP for all times $t > t_p$,
if there exists an operator ${\bm R} = {\bm R}^\dagger$ that
\begin{enumerate}[label=(\roman*)]
\item fulfils the operator inequality $\bm{\mathcal L}^\dagger \bullet \bm{R} + \bm{R} \bullet \bm{\mathcal L} \le 0$,
\label{cond:monotonicity}
\item is ``normalized'' as $\langle\bm{\Lambda}(t_p), (\bm{R}-\bm{S})\bullet\bm{\Lambda}(t_p)\rangle=0$,
\label{cond:normalization}
\item is such that  $\bm{R}-\bm{S}$ is positive semi-definite.
\label{cond:consistency}
\end{enumerate} 

To prove this statement, we employ \ref{cond:normalization} to reformulate \eqref{eq:representation_det_chi} to
\be
\label{eq:representation_det_chi_new}
e_h(\chi(t))=
e_h(\chi(t_p))
+\langle\bm{\Lambda}(t_p),{\bm R}\bullet \bm{\Lambda}(t_p)\rangle 
- \langle\bm{\Lambda}(t),{\bm S}\bullet \bm{\Lambda}(t)\rangle\ .
\ee
Since $e_h(\chi(t_p))$ is strictly positive by assumption, this implies the inequality
\be
\label{eq:estim}
e_h(\chi(t))>\langle\bm{\Lambda}(t_p), {\bm R}\bullet \bm{\Lambda}(t_p)\rangle
-\langle\bm{\Lambda}(t), {\bm S}\bullet \bm{\Lambda}(t)\rangle\ .
\ee
Condition \ref{cond:consistency} allows to bound this further to
\be
\label{eq:time_deriv}
e_h(\chi(t))> \langle\bm{\Lambda}(t_p), {\bm R}\bullet \bm{\Lambda}(t_p)\rangle
-\langle \bm{\Lambda}(t), {\bm R}\bullet \bm{\Lambda}(t) \rangle\ .
\ee
Due to \ref{cond:monotonicity} the time derivative of $\langle\bm{\Lambda}(t), {\bm R}\bullet\bm{\Lambda}(t)\rangle$ is always non-positive
\be
\partial_t\langle\bm{\Lambda}(t), {\bm R} \bullet \bm{\Lambda}(t)\rangle =  \langle\bm{\Lambda} (t), (\bm{\mathcal L}^\dagger \bullet{\bm R}+{\bm R}\bullet\bm{\mathcal L})\bm{\Lambda} (t)\rangle \le 0,
\ee
so that the right-hand-side of \eqref{eq:time_deriv} grows monotonically, which implies that $e_h(\chi(t))>0$ for $t > t_p > 0$.

Whereas $\chi(t_p)$ is (by assumption) positive semi-definite and of rank $h$, the ultimate goal is to show complete positivity for all times $t \ge 0$ and to also take into account the initial condition $e_h(\chi(0)) = 0$.
To this end it is necessary to verify that that all non-trivial eigenvalues of $\chi(t)$ become positive in leading order in $t$.
If this is the case, and $\langle\bm{\Lambda}(t), {\bm R}\bullet\bm{\Lambda}(t)\rangle$ is not strictly constant
(i.e.\ it decreases in leading order in $t$ following  \ref{cond:monotonicity}),
then \ref{cond:normalization} can be replaced by 
\begin{enumerate}[label=(\roman*')]
\setcounter{enumi}{1}
\item  $\langle\bm{\Lambda}(0), (\bm{R}-\bm{S})\bullet\bm{\Lambda}(0)\rangle = 0$
\label{condition:normalization_tp_equals_zero} 
\end{enumerate}
and \ref{cond:monotonicity}, \ref{condition:normalization_tp_equals_zero}, and  \ref{cond:consistency} verify CP dynamics for all $t \ge 0$.

To prove the latter statement, we first find $t_p > 0$ such that $\chi(t_p) \ge 0$ and $e_h(\chi(t_p))>0$, as well as $\langle\bm{\Lambda}(t_p), \bm{R}\bullet\bm{\Lambda}(t_p)\rangle < \langle\bm{\Lambda}(0),\bm{R}\bullet\bm{\Lambda}(0)\rangle$, and $\Lambda_1(t)$ is CP for all $0 \le t \le t_p$.
To show complete positivity for $t > t_p$, we rewrite \eqref{eq:representation_det_chi} trivially as
\be
\eqalign{
e_h(\chi(t))&=
\langle\bm{\Lambda}(0),\bm{S}\bullet \bm{\Lambda}(0)\rangle
-\langle\bm{\Lambda}(t_p), \bm{R}\bullet \bm{\Lambda}(t_p)\rangle\cr
&+\langle\bm{\Lambda}(t_p), {\bm R}\bullet \bm{\Lambda}(t_p) \rangle
- \langle\bm{\Lambda}(t), {\bm S}\, \bm{\Lambda}(t)\rangle\ .
}
\ee
Following \ref{condition:normalization_tp_equals_zero} this implies
\be
\eqalign{
e_h(\chi(t))&=
\langle\bm{\Lambda}(0), \bm{R}\bullet \bm{\Lambda}(0)\rangle
-\langle\bm{\Lambda}(t_p), \bm{R}\bullet \bm{\Lambda}(t_p)\rangle\\ 
&+\langle\bm{\Lambda}(t_p), \bm{R}\bullet \bm{\Lambda}(t_p) \rangle
- \langle\bm{\Lambda}(t), \bm{S}\bullet \bm{\Lambda}(t)\rangle\ ,
}
\ee
where the first two terms satisfy
$\langle\bm{\Lambda}(0), \bm{R}\bullet \bm{\Lambda}(0)\rangle
-\langle\bm{\Lambda}(t_p), \bm{R}\bullet \bm{\Lambda}(t_p)\rangle>0$, such that
\be
e_h(\chi(t))>
\langle\bm{\Lambda}(t_p), \bm{R}\, \bm{\Lambda}(t_p)  \rangle
- \langle\bm{\Lambda}(t), \bm{S}\, \bm{\Lambda}(t)\rangle\ .\nonumber
\ee
This proves strict positivity of $e_h(\chi(t))$ as shown above in \eqref{eq:estim}.

\subsection{Geometric interpretation}
\label{sec:geometric_interpretation}
\begin{figure}[t]
\centering
\includegraphics[scale=.25]{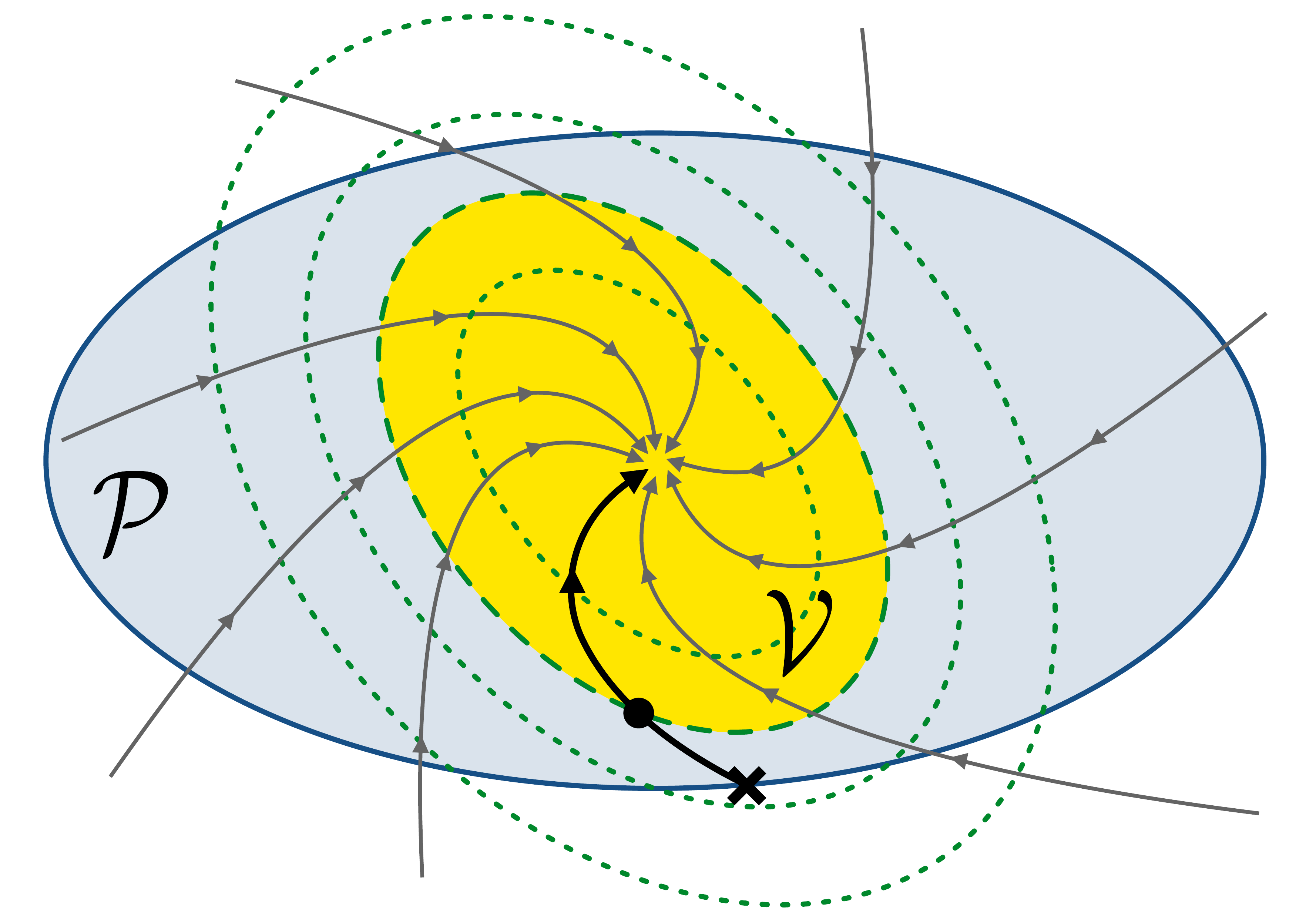}
\caption{A schematic representation of the set $\mathcal{P}$ (blue and yellow) of extended maps $\bm{\Gamma}=\sum_i\Gamma_i\otimes\ket{i}$ with CP $\Gamma_1$ is depicted.
The arrows show the dynamics according to \eqref{eq:maphierarchy}.
The thick trajectory corresponds to the dynamics that starts at $\bm{\Lambda}(0)$ (black cross) and passes through the point $\bm{\Lambda}(t_p)$ (black circle). Equipotential lines of the function ${\cal F}(\bm{\Lambda})=\langle\bm{\Lambda},\bm{R}\bullet\bm{\Lambda}\rangle$ are depicted in green (dotted).
The line that contains $\bm{\Lambda}(t_p)$ is depicted with long dashing and confines the set ${\cal V}$ (yellow).
If ${\cal V}$ is contained in ${\cal P}$ (as depicted), and ${\cal F}(\bm{\Lambda}(t))$ decreases monotonically, CP dynamics is verified for $t > t_p$.
} 
\label{fig:concept}
\end{figure}
The proposed procedure can be intuitively understood in geometrical terms:
Just like normalized, Hermitian operators are quantum states (i.e.\  positive semi-definite operators) of a two-level system if and only if they are represented by points inside the Bloch sphere, also valid extended dynamical maps form a geometrical object which we denote by $\cal P$.
Although the dimension of ${\cal P}$ is substantially larger than the dimension of the Bloch sphere, and its shape will typically be more complicated, this underlying geometry permits an immediate understanding of our approach.
The schematic representation of \fref{fig:concept} depicts all the points within ${\cal P}$ in blue and yellow.
In turn, all points outside this area correspond to invalid extended maps with a system component that is not CP.
The dynamics induced by \eqref{eq:maphierarchy} is indicated by arrows.

The goal is to find conditions which guarantee that, for a given initial condition $\bm{\Lambda}(t_p)$ within ${\cal P}$ (indicated by a black circle in \fref{fig:concept}), the dynamics is such that its trajectory will never leave ${\cal P}$.
As this can typically not be shown directly by means of $\cal P$, one strives for the identification of a region $\cal V$, such that
\vbox{%
\begin{itemize}
\item $\cal V$ contains $\bm{\Lambda}(t)$ for all times $t > t_p$
\item $\cal V$ lies inside $\cal P$.
\end{itemize}
}
This region $\cal V$ shall be confined by the equipotential line of a function ${\cal F}(\bm{\Lambda}) \equiv \langle\bm{\Lambda}, \bm{R}\bullet\bm{\Lambda}\rangle$ (\fref{fig:concept} depicts some equipotential lines of $\cal F$ in green (dotted/dashed)) that is determined by $\bm{R}$.
More precisely, $\cal V$ is given by all extended maps $\bm{\Gamma}=\sum_i \Gamma_i\otimes\ket{i}$ for which ${\cal F}(\bm{\Gamma})<{\cal F}(\bm{\Lambda}(t_p))$.
When the coordinate system is chosen such that the asymptotic extended map $\bm{\Lambda}(t \rightarrow \infty)$ lies in its origin, then ${\cal F}(\bm{\Lambda}(t))$ can be considered a distance between $\bm{\Lambda}(t)$ and the asymptotic map around which the object $\cal V$ is centred.

According to \ref{cond:monotonicity}, the function ${\cal F}(\bm{\Lambda}(t))$ decreases monotonically under the dynamics induced by \eqref{eq:hierarchy}, for why ${\cal F}(\bm{\Lambda}(t))$ can never exceed the value ${\cal F}(\bm{\Lambda}(t_p))$ for $t > t_p$.
The geometric implication of \ref{cond:monotonicity} is that $\bm{\Lambda}(t)$ will never leave $\cal V$.
Condition \ref{cond:normalization} and \ref{cond:consistency}, in turn, manifest a relation between $\cal P$ and $\cal V$ such that $\cal P$ contains $\cal V$.
This implies that and all extended maps inside $\cal V$ are valid.

The initial extended map $\bm{\Lambda}(0)$ (indicated by a black cross in \fref{fig:concept}) lies on the boundary of $\cal P$ (due to $\chi(0)$ having rank one) for why $\cal V$ and $\cal P$ become tangential in the point $\bm{\Lambda}(0)$ when $t_p$ becomes infinitesimally small.

\subsection{Simplification of condition \ref{condition:normalization_tp_equals_zero} and \ref{cond:consistency}}
\label{sec:simplification_of_condition_2_and_3}

The construction of $\bm R$ is typically most challenging when CP dynamics shall be proven for $t \ge 0$ (i.e.\ when $\cal V$ and $\cal P$ become tangential in $\bm{\Lambda}(0)$).
A particularly convenient case is, however, given when the $\bm S$ is of rank one, which, in geometrical terms, means that the valid region $\cal P$ is confined by two parallel planes.
The construction of $\bm{R} = \bm{R}^\dagger$ that proves complete positivity for $t \ge 0$ can in this case be simplified as the condition \ref{cond:consistency} can be partially replaced by a set of equality conditions, which effectively decrease the number of degrees of freedom in $\bm{R}$ that have to be determined.
This is particularly helpful for an analytical construction of the latter.

More specifically, the inequality ${\bm R}-{\bm S} \ge 0$ is satisfied (i.e.\ \ref{cond:consistency} holds true) if $\bm R$ is positive semi-definite, normalized as in \ref{condition:normalization_tp_equals_zero}, and satisfies
\be
\frac{\partial}{\partial \bm{K}^{(i)}} \langle\bm{\Gamma}, \bm{R}\bullet \bm{\Gamma}\rangle|_{\bm{\Gamma}=\bm{\Lambda}(0)} = 0
\label{eq:linear_conditions_on_R}
\ee
for a basis $\lbrace\bm{K}^{(i)}\rbrace$ of the kernel of $\bm S$.
\Eref{eq:linear_conditions_on_R} is a set of linear constraints that reduces the number of degrees of freedom in $\bm R$.

To understand why this reformulation of \ref{cond:consistency} is possible, we emphasize that $\bm{\Lambda}(0)$ lies on the boundary of $\cal P$ and can thus not be from the kernel of the operator $\bm S$.
That means that $\bm{\Lambda}(0)$ and $\lbrace \bm{K}^{(i)} \rbrace$ are linearly independent and form a complete set (because $\bm S$ is of rank one) such that every extended map $\bm v$ can be decomposed as $\bm{v} = a \bm{\Lambda}(0) + b \bm{K}$ with $\bm K$ being from the span of $\{\bm{K}^{(i)}\}$ and $a,b \in \mathbb{C}$.
Without loss of generality, it can be assumed\footnote{If that is not the case, then one can consider $b' \equiv b+\Im(a^*b)/\Im(a)$ and $\bm{K'} \equiv  (1 - \Im(a^*b)/[b'\Im(a)])\bm{K} \in \textrm{ker}(\bm{R})$ such that $\bm{V} = a\bm{\Lambda}(0) + b'\bm{K'}$ and $\Im(a^*b')=0$} that $a^*b \in \mathbb{R}$.
With this decomposition, we obtain
\be
\langle\bm{v}, (\bm{R}-\bm{S})\bullet \bm{v}\rangle = a^*b(\langle\bm{\Lambda}(0), \bm{R}\bullet \bm{K} \rangle + \langle\bm{K}, \bm{R}\bullet \bm{\Lambda}(0)\rangle) + b^2 \langle\bm{K},\bm{R}\bullet \bm{K}\rangle,
\label{eq:proof_positivity}
\ee
where \ref{condition:normalization_tp_equals_zero} has been employed.
From \eqref{eq:linear_conditions_on_R}, one can, however, deduce
\begin{eqnarray}
\fl
\langle\bm{\Lambda}(0), \bm{R} \bullet \bm{K}\rangle + \langle\bm{K}, \bm{R}\bullet \bm{\Lambda}(0)\rangle &=
\lim_{\epsilon\rightarrow 0} \frac{1}{\epsilon}[\langle\bm{\Lambda}(0) + \epsilon \bm{K}, \bm{R}\bullet (\bm{\Lambda}(0) + \epsilon \bm{K})\rangle - \langle\bm{\Lambda}(0), \bm{R}\bullet \bm{\Lambda}(0)\rangle]\nonumber\\ 
&= \frac{\partial}{\partial \bm{K}} (\bm{\Gamma}, \bm{R}\bullet \bm{\Gamma})|_{\bm{\Gamma}=\bm{\Lambda}(0)}
=0\ ,
\end{eqnarray}
which, together with $\bm{R} \ge 0$, implies non-negativity of \eqref{eq:proof_positivity} and thus positive semi-definiteness of $\bm{R}-\bm{S}$.

With the framework derived hitherto, we are now in the position to analyse exemplary equations of motion that permit to construct $\bm{R}$ by purely analytic means.
Subsequently, in \sref{sec:transformation_towards_simplified_geometries} we will expand the framework of constructing $\bm{R}$ towards numerical techniques that will permit to treat more complex cases as exemplified in \sref{sec:example_non-markovian_bath_of_finite_temperature} and \ref{sec:example_the_spin-boson_model}

\subsection{Examples}
\label{sec:introductory_examples}
\subsubsection{The damped Jaynes-Cummings model}
\label{sec:damped_jaynes_cummings}
An instructive example of a HEOM is given by the damped resonant Jaynes-Cummings model, which describes the dynamics of a two-level system inside a leaky cavity.
The electro-magnetic cavity field is characterized by a Lorentzian spectral density function that is centred around the transition frequency of the two-level system.
The spectral width of the cavity field is denoted by $\zeta$, whereas $\gamma$ labels the coupling strength of the two-level system and the cavity field.
After tracing out the electromagnetic field, this model permits an exact description of the (generally non-Markovian) dynamics in terms of a HEOM, which reads
\vbox{%
\numparts
\begin{equationarray}{rcr@{}c@{}lr@{}c@{}lr@{}c@{}l}
\vphantom{\sum_i^j}
\dot{\varrho}_1(t) &=&&&&\zeta\,&\varrho_2(t)&&&&\label{eq:jaynes_cummings_hierarchy_a}\\
\vphantom{\sum_i^j}
\dot{\varrho}_2(t) &=& \gamma\,{\cal D}_{-}\,&\varrho_1(t)&& +\, \zeta\sum_{i =x,y,z} \,\sigma_i &\varrho_2(t)&\sigma_i^\dagger\ &+\,  \zeta\,&\varrho_3(t)&\label{eq:jaynes_cummings_hierarchy_b}\\
\vphantom{\sum_i^j}
\dot{\varrho}_3(t) &=&&&&\frac{\gamma}{2}\sum_{i =x,y} \sigma_i & \varrho_2(t) & \sigma_i^\dagger &-\,2 \zeta \,\sigma_z^\dagger& \varrho_3(t)&\sigma_z\label{eq:jaynes_cummings_hierarchy_c}
\end{equationarray}
\endnumparts
}
with ${\cal D}_{-}\,\varrho_1 = \sigma_-\varrho_1\sigma_+-\{\sigma_+\sigma_-,\varrho_1\}/2$ and $\sigma_\pm = (\sigma_x \pm i\sigma_y)/2$.
The system state is denoted by $\varrho_1(t)$, whereas the two auxiliary operators $\varrho_2(t)$ and $\varrho_3(t)$ encode information about the environment and system-environment correlation.
The introduction of a general framework by means of which this HEOM has be obtained is given in \ref{sec:targeted_solution} and remarks that explicitly concern the construction of \eqref{eq:jaynes_cummings_hierarchy_a} - \eqref{eq:jaynes_cummings_hierarchy_c} are found in \ref{sec:jaynes_cummings}.

To turn this HEOM equation into an equation of motion for the dynamical maps $\Lambda_1(t)$ to $\Lambda_3(t)$, one writes \eqref{eq:jaynes_cummings_hierarchy_a}-\eqref{eq:jaynes_cummings_hierarchy_c} in the Bloch basis. 
The Bloch representation of the extended state is then given by a $12$-dimensional vector,
and the $12\times 12$-dimensional Bloch representation for the generator $\bm{\mathcal{L}}$,
is also the generator for the dynamics of the $12\times 4$-dimensional Bloch representation of the extended dynamical map.
The latter is initialized by $\Lambda_1(0) = \mathbb{I}$ and $\Lambda_i(0)=\mathbb{O}$ for $i \ge 2$, where $\mathbb{I}$ and $\mathbb{O}$ are the identity and the null map, respectively.
This corresponds to the initial condition $\varrho_2(0)=\varrho_3(0)=\mathbb{O}$.
It has been algebraically verified that the coefficient matrix $\chi(t)$, which characterizes the system map $\Lambda_1$ (see \eqref{eq:Kraus} with $\mu_i = \sigma_i$), is always of the form
\be
\label{eq:jaynes_cummings_strucutre_of_chi}
\chi(t) = \frac{1}{4}
\left[
\begin{array}{cccc}
 (\lambda_1(t)+1)^2 & 0 & 0 & \lambda_1^2(t)-1 \\
 0 & 1-\lambda_1^2(t) & i - i \lambda_1^2(t) & 0 \\
 0 & i\lambda_1^2(t)-i & 1-\lambda_1^2(t) & 0 \\
\lambda_1^2(t)-1 & 0 & 0 & (\lambda_1(t)-1)^2
\end{array}
\right]\ ,
\ee
where the only time-dependent degree of freedom $\lambda_1(t)$ is a coordinate of the system map $\Lambda_1(t)$.
Thus, the condition $\chi(t) \ge 0$, i.e.\ complete positivity of the system map $\Lambda_1(t)$, does only depend on $\lambda_1(t)$ and all other degrees of freedom of $\Lambda_1(t)$ can safely be ignored.
The dynamics of $\lambda_1(t)$ is obtained from the differential equation $\partial_t\bm{\Lambda}(t) = \bm{\mathcal L}\bullet\bm{\Lambda}(t)$ and reads $\partial_t\vec\lambda(t) = l\,\vec\lambda(t)$, where $\vec\lambda(t) = [\lambda_1(t),\lambda_2(t)]$ is a two-dimensional real-valued vector and
\be
l = 
\left[
\begin{array}{cc}
0 & \zeta \\
-\frac{\gamma}{2} & -\zeta
\end{array}
\right]\ .
\ee
The dynamical variable $\lambda_1(t)$ is a coordinate of $\Lambda_1(t)$ and the initial condition $\bm{\Lambda}(0)$ corresponds to $\vec\lambda(0) = [1,0]$.
Strict positivity of all non-trivial eigenvalues of $\chi(t)$ (two of them vanish constantly) in the first non-vanishing time-step is given when $\gamma\zeta>0$.
When this product is positive, we can replace the normalization condition \ref{cond:normalization} by \ref{condition:normalization_tp_equals_zero}.

\begin{figure}[t!]
\centering
\includegraphics[scale=.55]{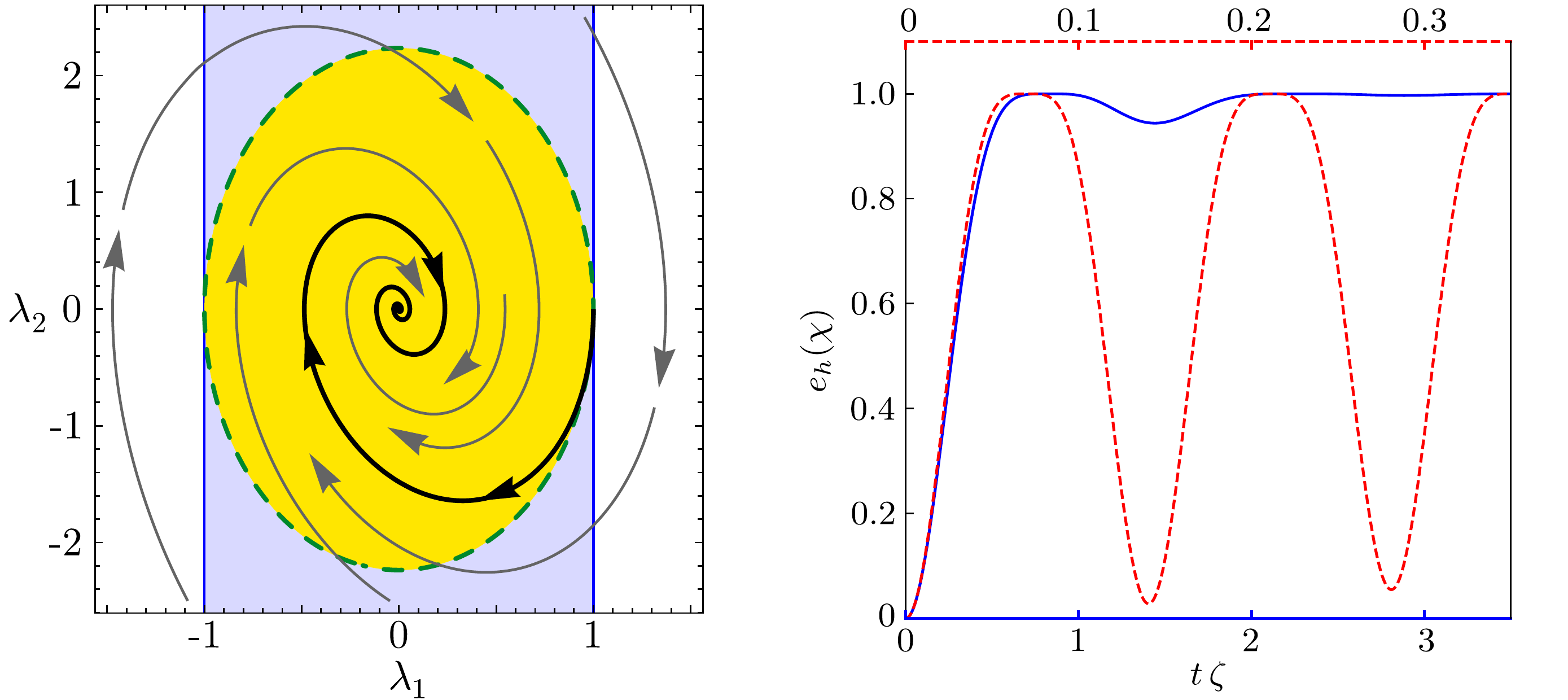}
\caption{Figure (a) depicts a representation of the set $\mathcal{P}$ (blue/yellow, solid) of extended dynamical maps with a system component that is CP.
The green (dashed) equipotential line of the monotone ${\cal F}$ contains the initial condition $\vec\lambda(0)$ (black cross) and bounds the set ${\cal V}$ (represented in yellow). The dynamics induced by \eqref{eq:jaynes_cummings_hierarchy_a}-\eqref{eq:jaynes_cummings_hierarchy_c}, is depicted by vectors (grey/black) and the trajectory for the initial condition $\vec\lambda(0)$ is indicated in black (thick line). The set ${\cal V}$ lies completely inside ${\cal P}$ and as ${\cal F}$ decreases monotonically, this proves complete positivity. Figure (b) shows $e_h(\chi)$ as a function of time and gives an idea of what values the smallest non-trivial eigenvalues of $\chi(t)$ can take. Blue (solid) corresponds to $\gamma=10\zeta$ (same parameters as in (a)), whereas red (dashed) represents $\gamma=10^5\zeta$. Both cases are covered by the conditions derived in \sref{sec:damped_jaynes_cummings}.}
\label{fig:jaynes_cummings}
\end{figure}

Due to the structure of $\chi(t)$ in \eqref{eq:jaynes_cummings_strucutre_of_chi}, its highest and the second-highest elementary symmetric polynomials $e_4(\chi(t))$ and $e_3(\chi(t))$
 are constantly equal to zero.
The complete positivity of the system map $\Lambda_1(t)$ thus depends on the highest non-vanishing polynomial $e_h(\chi) = e_2(\chi) = (\textrm{tr}(\chi)^2 - \textrm{tr}(\chi^2))/2= (1-\lambda_1^4)/4$ which is positive if and only if $|\lambda_1(t)|<1$.
This condition is expressed as $e_h(\chi(t)) = \vec\lambda^\dagger(0)\,s\,\vec\lambda(0) - \vec\lambda^\dagger(t)\,s\,\vec\lambda(t) > 0$ with a matrix $s = \textrm{diag}[1,0]/4$ that characterizes the valid region $\cal P$.
The situation is depicted in \fref{fig:jaynes_cummings}(a).

To prove strict positivity of $e_h(\chi(t))$ for all times $t>0$, we strive for a matrix $r$ (the representation of $\bm R$ in the space of $\vec\lambda(t)$)  that satisfies $l^\dagger\,r + r\,l \le 0$ (condition \ref{cond:monotonicity}), $\vec\lambda^\dagger(0)\,(r-s)\,\vec\lambda(0) = 0$ (condition \ref{condition:normalization_tp_equals_zero}), and $r-s \ge 0$ (condition \ref{cond:consistency}).
Condition~\ref{condition:normalization_tp_equals_zero} and \ref{cond:consistency} hold true, when $r$ is parametrized by $r=\textrm{diag}(1,r_{22})/4$ with $r_{22} \ge 0$.
Condition~\ref{cond:monotonicity}, in turn, is satisfied if $q \equiv l^\dagger\,r + r\,l \le 0$, which is the case if and only if trace and determinant of $-q$ are non-negative.
The latter is given as $\det(-q) = - (\gamma r_{22}-2\zeta)^2/4$ for why we have to choose $r_{22} = 2\zeta/\gamma$ and with $\textrm{tr}(-q) = 2 r_{22} \zeta =4\zeta^2/\gamma$ the conditions $q \le 0$ and $r_{22} \ge 0$ are equivalent to $\gamma \ge 0$ and $\zeta \ge 0$.
Together with $\gamma\zeta>0$ this proves that positivity of both parameters $\gamma$ and $\lambda$ is sufficient for $e_h(\chi(t))>0$ and implies CP dynamics.
Even though eigenvalues can become very small (see \fref{fig:jaynes_cummings}(b) for an example), the conditions clearly assert that no eigenvalue of $\chi(t)$ vanishes.

\subsubsection{Reviving coherences: a HEOM with two levels}
\label{sec:reviving_coherences_a_heom_with_two_levels}
Another model for a non-monotonic decay of phase coherence can be formulated in terms of the hierarchy equation (see \ref{sec:appendix_two_level_reviving_coherences} for an explicit derivation)\\
\vbox{%
\numparts
\begin{equationarray}{rcr@{}c@{}lr@{}c@{}l}
\dot{\varrho}_1(t) &=&\frac{\gamma_1}{2}D_z&\varrho_1(t)&&+\,\omega & \varrho_2(t) \label{eq:hierarchyExample2da}&\\
\dot{\varrho}_2(t) &=&\alpha\omega D_z&\varrho_1(t)&&+\, \gamma_2 \sigma_z & \varrho_2(t)&\sigma_z
\label{eq:hierarchyExample2db}
\end{equationarray}
\endnumparts
}
with the Pauli operator $\sigma_z$ and the dephasing Lindbladian $D_z\varrho = \sigma_z\varrho\sigma_z - \varrho$.
For $\alpha=0$ and the initial condition $\varrho_2(0)=0$, this hierarchy equation describes purely Markovian dynamics with exponentially decaying coherence.
For non-vanishing values of $\alpha$, however, the dynamics deviates from a mere exponential decay and can give rise to genuine non-Markovian revivals. 
The initial condition $\varrho_2(0)$ governs the infinitesimal behaviour of $\varrho_1(t)$ at time $t=0$. A generic choice is $\varrho_2(0)=0$ although other choices are possible as we will demonstrate later.

In the Bloch basis, the extended state $\bm{\varrho}$ can be expressed in terms of an eight-dimensional vector and the according representation of the generator $\bm{\mathcal L}$ reads
\be
\label{eq:heom_2d_equation_of_motion}
\left[
\begin{array}{cccc|cccc}
   0 & 0 & 0 & 0 & \omega & 0 & 0 & 0 \\
   0 & -\gamma_1  & 0 & 0 & 0 & \omega & 0 & 0 \\
   0 & 0 & -\gamma_1  & 0 & 0 &  0 & \omega & 0 \\
   0 & 0 & 0 & 0 & 0 & 0 & 0 & \omega \\\hline
   0 & 0 & 0 & 0 & \gamma_2 & 0 & 0 & 0 \\
   0 & -2\alpha\omega  & 0 & 0 & 0 & -\gamma_2  & 0 & 0 \\
   0 & 0 & -2\alpha\omega & 0 & 0 & 0 & -\gamma_2  & 0 \\
   0 & 0 & 0 & 0 & 0 & 0 & 0 & \gamma_2  \\
\end{array}
\right]\ .
\ee
With the initial conditions $\Lambda_1(0) = \mathbb{I}$, and $\Lambda_2(0)=\mathbb{O}$ and the structure of $\bm{\mathcal L}$ one can show by purely algebraic means that the Bloch representations for the extended dynamical map $\bm{\Lambda}(t)$ reads 
 $\textrm{diag}[1,\lambda_1(t),\lambda_1(t),1]\otimes\ket{1}+\textrm{diag}[0,\lambda_2(t),\lambda_2(t),0]\otimes\ket{2}$
and depends only on the two scalar parameters $\lambda_1(t)$ and $\lambda_2(t)$.
With this form of the dynamical map $\Lambda_1(t)$, the corresponding matrix $\chi(t)$ is given by $\chi(t)=\textrm{diag}[1+\lambda_1(t),0,0,1-\lambda_1(t)]/2$.
In particular, the maximal rank of $\chi(t)$ is $h=2$ and complete positivity of $\Lambda_1(t)$ is given for $|\lambda_1(t)|<1$, or, equivalently, $e_h(\chi(t)) = \vec\lambda^\dagger(0)\,s\,\vec\lambda(0) - \vec\lambda^\dagger(t)\,s\,\vec\lambda(t) >0$ with $s=\textrm{diag}[1,0]$.
Strict positivity of all non-trivial eigenvalues of $\chi(t)$ for infinitesimal times is given when either $\gamma_1 > 0$, or $\gamma_1=0$ and $\omega,\alpha >0$ hold true.

The procedure of proving complete positivity is now analogous to the previous case:
As $\lambda_1(t)$ and $\lambda_2(t)$ are the only relevant dynamical variables, the equation of motion for the extended dynamical map can be reduced to $\partial_t \vec\lambda(t) = l\,\vec\lambda(t)$ with $\vec\lambda(t)=[\lambda_1(t), \lambda_2(t)]$, the initial condition $\vec\lambda(0)=[1,0]$ and the generator
\be
\label{eq:reduced_differential_operator_revivingCoherence_2d}
l = 
\left[
\begin{array}{cc}
-\gamma_1 & \omega \\ 
-2\alpha\omega & -\gamma_2
\end{array}\right]\ .
\ee

\begin{figure}[t]
\centering
\includegraphics[scale=.8]{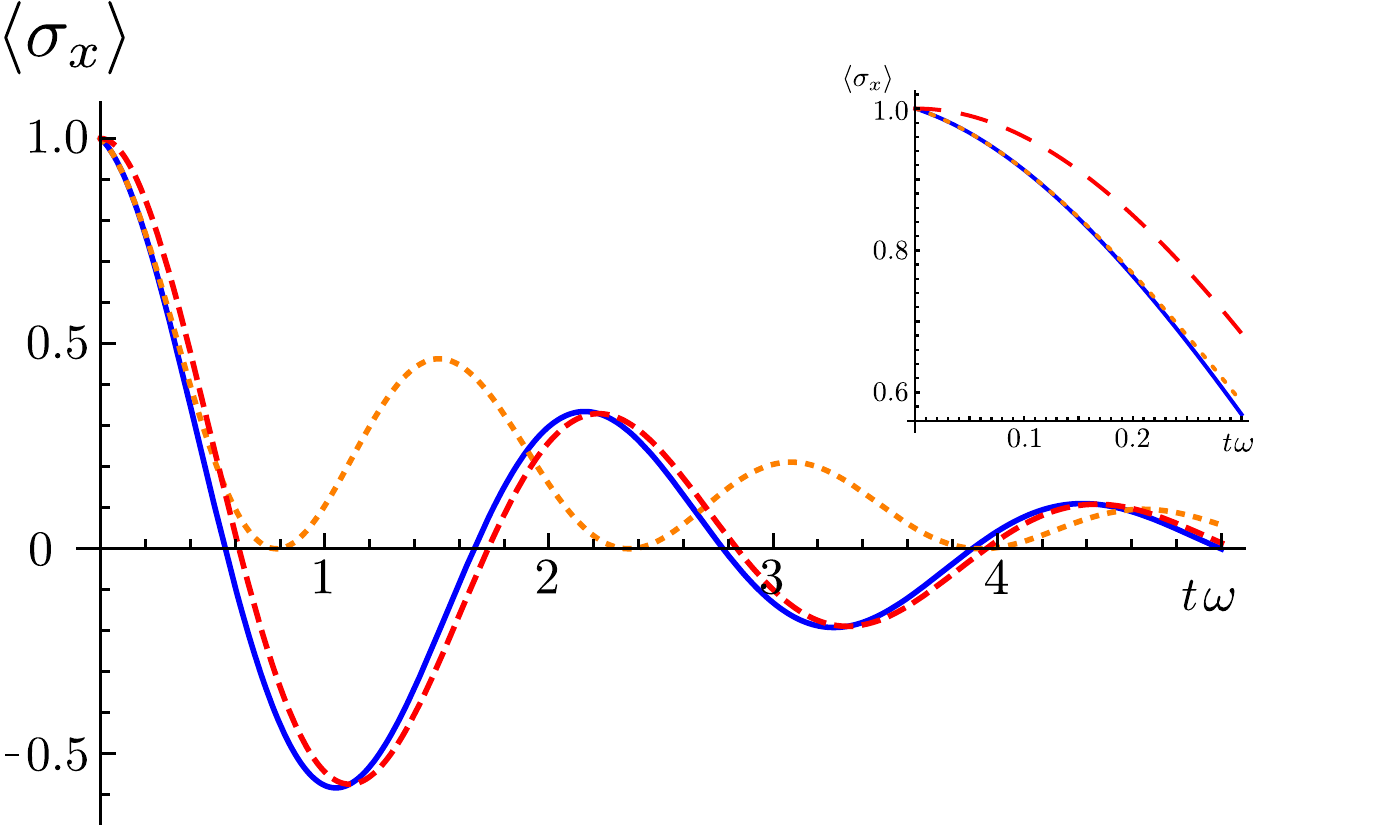}
\caption{Time evolution of the coherence $\tr(\sigma_x\varrho_1)$ for an initial state $\varrho_1(0)=\prj{+}$ with $\ket{+}=(\ket{0}+\ket{1})/\sqrt{2}$,
evolving according to \eqref{eq:hierarchyExample2da}, \eqref{eq:hierarchyExample2db}, and its extension \eqref{eq:stateHierarchyExample3d}. The dynamics of \eqref{eq:hierarchyExample2da} and \eqref{eq:hierarchyExample2db} with initial conditions $\varrho_2(0)=0$ and $\dot\varrho_1(0)=0$ is depicted in blue (solid) and red (long dashing), respectively ($\gamma_1=\gamma_2=1/2\,\omega,\ \alpha=4$). The orange (short dashing) line represents the dynamics induced by introducing the additional level in \eqref{eq:stateHierarchyExample3d} with $\varrho_2(0)=\varrho_3(0)=0$  ($\beta=8,\ \gamma=1/2\,\omega$).
}
\label{fig:reviving_cohrences_examples}
\end{figure}

As already seen for the Jaynes-Cummings model, the operator $\bm R$ is characterized by a $2 \times 2$-dimensional matrix $r$ that is parametrized by $r = \textrm{diag}[1,r_{22}]/4$ with $r_{22} \ge 0$ in order to satisfy the conditions \ref{condition:normalization_tp_equals_zero} and \ref{cond:consistency}.
The monotonicity condition \ref{cond:monotonicity} holds true if $l^\dagger\,r + r \, l \le 0$, and a good choice for $r_{22}$ (the choice is not unique) is given by $r_{22} = (\alpha\omega^2 + \gamma_1\gamma_2)/(2\alpha^2\omega^2)$.
In this case, monotonicity and $r_{22} \ge 0$ are equivalent to $\gamma_1 \ge 0$, $\gamma_2 \ge 0$ and  $2\alpha+\gamma_1\gamma_2\ge 0$ such that these three conditions together with $\varrho_2(0)=\mathbb{O}$ and strict positivity of $\chi(t)$ for short times are sufficient for the  hierarchy in \eqref{eq:hierarchyExample2da} and \eqref{eq:hierarchyExample2db} to induce CP dynamics.
An example of a process that satisfies these conditions is given in \fref{fig:reviving_cohrences_examples}.

It is well established that the initially linear decay of coherence, that is depicted in the inset of \fref{fig:reviving_cohrences_examples}, is an artefact of the Markov approximation.
In the framework of hierarchy equations, it is straight-forward to remove this artefact by choosing the initial condition $\varrho_2(0)=-{\mathcal L}_{11}\,\varrho_1(0)$, which corresponds to $\vec\lambda(0) = [1,\gamma_1/\omega]$ and yields $\dot\varrho_1(0) = 0$.
\Fref{fig:reviving_cohrences_examples} shows the difference of the time evolution in the first time-steps.

The examination of complete positivity requires a modification of the matrix $r$ in order to still comply with \ref{condition:normalization_tp_equals_zero} due to the change of the initial condition $\vec\lambda(0)$.
To this end, we parametrize $r$ by means of the Cholesky decomposition such that it is positive semi-definite and $\vec\lambda(0)\,(r-s)\vec\lambda(0) = 0$, which (due to the symmetry of $r$) reduces the number of degrees of freedom in $r$ to two.
The condition $r-s \ge 0$ is then incorporated into the parametrization of $r$ as shown in \sref{sec:simplification_of_condition_2_and_3}, which eliminates another degree of freedom.
With this parametrization, the two conditions \ref{condition:normalization_tp_equals_zero} and \ref{cond:consistency} are satisfied and all but one single degree of freedom of $r$ have been determined.
This last degree of freedom can for example be chosen such that $\det(l^\dagger\,r + r\,l) = 0$.
In this case one obtains
\be
r=\frac{1}{4 \eta}\left[
\begin{array}{cc}
 \gamma_1^2+\eta & -\omega\gamma_1 \\
 -\omega\gamma_1 & \omega^2 \\
\end{array}
\right]
\ee
with $\eta = 2\alpha\omega^2 +\gamma_1 \gamma_2$.
The monotonicity condition \ref{cond:monotonicity} and $r \ge 0$ are then equivalent to $\gamma_1 + \gamma_2 \ge 0$ and $2\alpha\omega^2 + \gamma_1\gamma_2 > 0$, which (together with the condition that all non-trivial eigenvalues of $\chi(t)$ become strictly positive for short times $t$) are thus sufficient conditions for CP dynamics.

\subsubsection{Reviving coherences: a HEOM with three levels}
The HEOM in \eqref{eq:hierarchyExample2da} and \eqref{eq:hierarchyExample2db} can be extended by adding a new auxiliary operator $\omega \varrho_3(t)$ to the right-hand side of \eqref{eq:hierarchyExample2db} and augmenting the equation of motion by (see \ref{sec:extension_of_lindblad} for a rigorous derivation of this equation)
\be
\label{eq:stateHierarchyExample3d}
\dot{\varrho}_3(t) = \beta\omega\,\sigma_z\,\varrho_2(t)\,\sigma_z^\dagger + \gamma_3\,\sigma_z\,\varrho_3(t)\,\sigma_z^\dagger\ .
\ee
Such HEOM gives rise to a broader variety of dynamical processes and rekindles the question of complete positivity.
In favour of a concise discussion, we reduce the number of free parameters by setting $\gamma_1 = \gamma_2 = \gamma_3 \equiv \gamma$.

The extended state $\bm{\varrho}(t)$ as well as the generator $\bm{\mathcal L}$ are again expressed in the Bloch basis.
In similarity to \eqref{eq:heom_2d_equation_of_motion}, this permits to express the HEOM $\partial_t\bm{\varrho}(t) = \bm{\mathcal L}\bullet\bm{\varrho}(t)$ in terms of a vector equation and the
transition from states to dynamical maps is made by replacing $\bm{\varrho}(t)$ by the extended dynamical map $\bm{\Lambda}(t)$, which is initialized by $\Lambda_1(0)=\mathbb{I}$ and $\Lambda_i(0)=\mathbb{O}$ for $i=2,3$.
In the Bloch representation, all dynamical maps $\Lambda_i(t)$ preserve the structure $\textrm{diag}[\delta_{i1},\lambda_i(t),\lambda_i(t),\delta_{i1}]$
 such that the dynamics of $\bm{\Lambda}(t)$ is completely characterized by the three-dimensional real-valued vector $\vec\lambda(t)$, which is initialized by $\vec\lambda(0)=[1,0,0]$ and evolves according to $\partial_t\vec\lambda(t) = l \,\vec\lambda(t)$ with
\be
\label{eq:reviging_coherence_3d_l}
l=
\left[
\begin{array}{ccc}
 -\gamma  & \omega & 0 \\
 -2\alpha\omega  & -\gamma  & \omega \\
 0 & -\beta\omega  & -\gamma  \\
\end{array}
\right]\ .
\ee
In fact, the question of complete positivity only depends on the two ratios $\tilde{\alpha}=\alpha\omega^2/\gamma^2$ and $\tilde{\beta}=\beta\omega^2/\gamma^2$,
as one can see in terms of the scaled variables $\tilde\lambda_i(t)=\lambda_i(t)\omega^i/\gamma^i$, whose equations of motion is described by $\tilde l$ that is obtained from \eqref{eq:reviging_coherence_3d_l} by multiplying all elements $l_{ij}$ with $(\omega/\gamma)^{i-j}$.
Since the structure of $\chi(t)$ does not change as compared to the HEOM with two levels, its highest non-trivial elementary symmetric polynomial $e_h = e_2$ is still correctly expressed by $e_h(\chi(t)) = \vec\lambda^\dagger(0)\,s\,\vec\lambda(0) - \vec\lambda^\dagger(t)\,s\,\vec\lambda(t)$ with $s = \textrm{diag}[1,0,0]/4$.
Again, a Cholesky decomposition of $r$ (such that $r \ge 0$) and the condition $\vec\lambda^\dagger(0)(r-s)\vec\lambda(0)=0$ permit to express $r-s \ge 0$ in terms of two linear equality constraints (see \sref{sec:simplification_of_condition_2_and_3}), which give rise to the parametrization 
\be
r = \frac{1}{4}\left[
\begin{array}{ccc}
 1 & 0 & 0 \\
 0 & a & c a \\
 0 & c a & a c^2+b \\
\end{array}
\right]\ 
\ee
with $a,b\ge 0$.
Considering the more interesting case of $\tilde\alpha,\tilde\beta \ne 0$, a very convenient choice for the three free parameters is given by
$a  =1/|2 \tilde\alpha |$, $b =1 / |2 \tilde\alpha \tilde\beta |$ and $c=0$.
This matrix satisfies the monotonicity condition $l^\dagger r + rl \le 0$ if and only if one of the following two conditions is satisfied:
\numparts
\begin{equationarray}{r@{}c@{}lcr@{}c@{}l@{}}
-1/2 \le &\, \tilde\alpha & \,<0 &\textrm{and}&-2\tilde\alpha-1 &\,\le\,& \tilde\beta \label{eq:reviving_coherence_3d_a}\\
& \tilde\alpha & \,> 0 &\textrm{and}&-1 &\,\le\,&  \tilde\beta\ .\label{eq:reviving_coherence_3d_b}
\end{equationarray}
\endnumparts
Consequently, any of the two conditions \eqref{eq:reviving_coherence_3d_a} or \eqref{eq:reviving_coherence_3d_b} ensures complete positivity when strict positivity of all non-trivial eigenvalues of $\chi(t)$ is given for short times.

\begin{figure}[t]
\centering
\includegraphics[scale=1.]{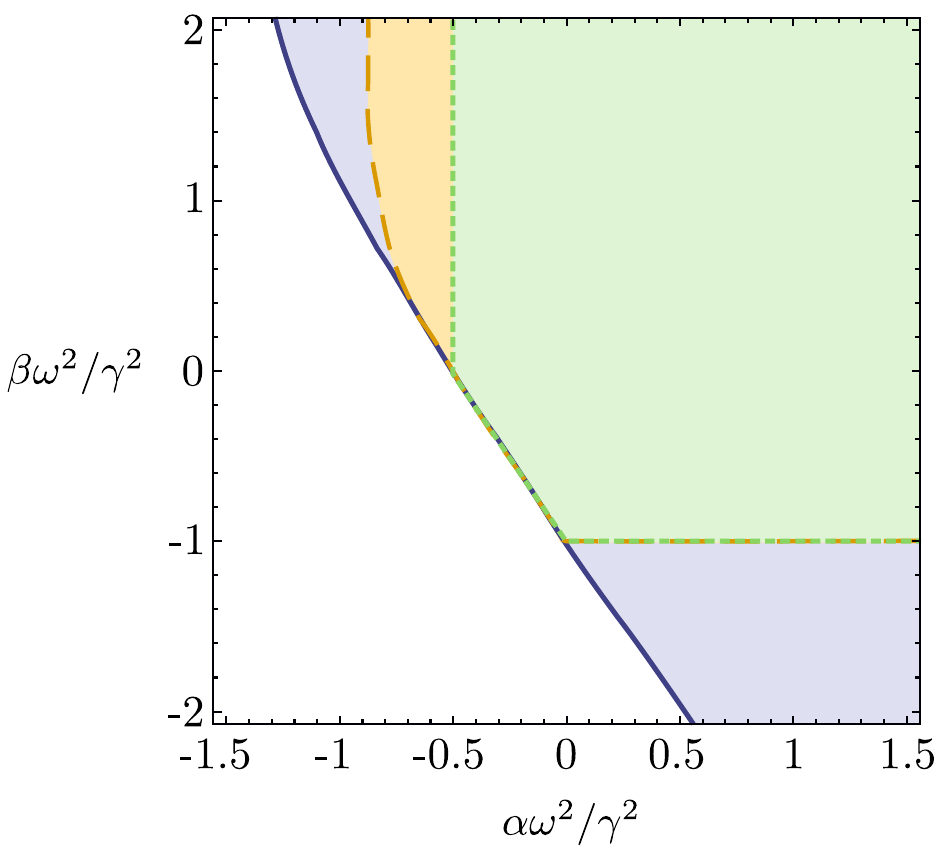}
\caption{Parameter set for which the system dynamics induced by the HEOM in \eqref{eq:hierarchyExample2da}, \eqref{eq:hierarchyExample2db} and \eqref{eq:stateHierarchyExample3d} is found to be CP.
Green (short-dashed) corresponds to purely analytic verification of complete positivity, and orange (long-dashed) is based on a numerical optimization of the operator $\bm R$.
White corresponds to violation of complete positivity identified through explicit solution of the equations of motion.
The blue region (solid), thus seems to correspond to CP dynamics, but can not be identified as such with the present framework.
}
\label{fig:compPlot}
\end{figure}

Although this is only one instance of how the matrix $r$ can be chosen and other choices (e.g.\ with $c \neq 0$) are possible, it can prove complete positivity for a rather large parameter regime as shown in \fref{fig:compPlot}. 
An individual numeric optimization over $r$ for each numerical instance of the HEOM can even slightly enlarge that parameter set.

\section{Numeric approaches to more complex problems}
\label{sec:transformation_towards_simplified_geometries}
\subsection{Transformation of the extended dynamical map}
\label{sec:transformation_of_the_extended_dynamical_map}
A central element of the procedure discussed in ~\sref{sec:monotone-based_conditions_for_complete_positivity} is to express the highest non-trivial elementary symmetric polynomial $e_h(\chi(t))$ in terms of a quadratic function in $\bm{\Lambda}(t)$ as shown in \eqref{eq:representation_det_chi}.
For the examples discussed in \sref{sec:introductory_examples}, such a representation has been obtained by anticipating the structure of the matrix $\chi(t)$ that characterizes potential solution to the initial value problem by purely algebraic means.
Because the initial condition $\Lambda_1(0)=\mathbb{I}$ and $\Lambda_i(0)=\mathbb{O}$ for $i \ge 2$ is fixed, it is for example often possible to prove that particular subspaces are never occupied.
To this end, one shows that the dimension of the space that is spanned by $\bm{\mathcal L}^{\bullet k}\bm{\Lambda}(0)$ with $k=0,\ldots,d$ and $d$ being the number of degrees of freedom in $\bm{\Lambda}(t)$ is strictly smaller than $d$.
When that is the case, then this indicates the existence of preserved quantities, which permit predictions about the form of $\chi(t)$ and can thus help to obtain \eqref{eq:representation_det_chi} (and to reduce the dimensionality of $\bm{\Lambda}(t)$).

To make the approach more universal and to permit $e_h(\chi(t))$ to be of a degree $m > 2$ in $\chi(t)$, one can apply a transformation which renders geometries that are always accessible to our technique.
To this end, we define a function
\be
\eqalign{
G[\chi(t)] &\equiv c^2 - (e_h(\chi(t))-c)^2 = 2c\, e_h(\chi(t)) - e_h^2(\chi(t)) \nonumber\\
& \le 2c\,e_h(\chi(t))\ ,
}
\ee
with a constant $c>0$ such that the strict positivity of $G[\chi(t)]$ is sufficient for $e_h(\chi(t))>0$.
Although $G[\chi(t)]$ is of a higher degree in $\bm{\Lambda}(t)$, considering the function $G[\chi(t)]$ rather than $e_h(\chi(t))$ has the big advantage that the prior can always be expressed in a form that is very similar to \eqref{eq:representation_det_chi}.
This is done by defining a new vector $\vec\Xi(t)$ whose first component reads $\Xi_1(t)=e_h(\chi(t))-c$.
Independently of the components $\Xi_i(t)$ with $i \ge 2$, this vector permits to express $G[\chi(t)]$ as
\be
\label{eq:representation_of_G}
G[\chi(t)] = \underbrace{\vec\Xi^\dagger(0)\,\tilde{S}\,\vec\Xi(0)}_{c^2} - \underbrace{\vec\Xi^\dagger(t)\,\tilde{S}\,\vec\Xi(t)}_{(e_h(\chi(t))-c)^2}\ ,
\ee
with a matrix $\tilde{S}$ that is of the simple form $\tilde{S}=\textrm{diag}[1,0,\ldots,0]$.
It is essential that $\vec\Xi(t)$ follows a linear differential equation $\partial_t \vec\Xi(t) = \tilde{\mathcal{L}}\,\vec\Xi(t)$ and as the elements of $\vec\Xi(t)$ are functions of $\chi(t)$ (and thus $\bm{\Lambda(t)}$), this differential equation needs to be derived from the HEOM.
Already the first component $\Xi_1(t)$, however, is a non-linear function in $\bm{\Lambda}(t)$ and its time-derivative does typically also involve terms that are non-linear in $\bm{\Lambda}(t)$. For $\vec\Xi(t)$ to anyway evolve according to a linear equation, one can now consider any non-linear contribution to be a new element of $\vec\Xi(t)$ and keep increasing the size of $\vec\Xi(t)$ until one has recovered a linear equation of motion $\partial_t\vec\Xi(t)=\tilde{\mathcal L}\,\vec\Xi(t)$ \footnote{A simple example would be the scalar equation of motion $\dot x(t)=\alpha x(t)$ and the non-linear coordinate transformation $y(t)=x^2(t)+x(t)$. The coordinate $y(t)$ satisfies $\dot y(t)=2\alpha y(t)-\alpha x(t)$; replacing $x(t)$ in terms of $y(t)$ would result in a non-linear equation of motion, but treating $x(t)$ as independent variable, permits to maintain linear equation of motion at the expense of increasing the number of independent variables.}.
The matrix $\tilde{\mathcal L}$ is constructed accordingly and eventually yields the desired equation of motion for $\vec\Xi(t)$.

In order to prove strict positivity of $G[\chi(t)]$, one follows the same procedure that has been introduced in \sref{sec:the_algebraic_framework} and seeks a matrix $\tilde{R} \ge 0$ that satisfies \ref{cond:monotonicity} to \ref{cond:consistency} (where $\bm{\mathcal L}$ and $\bm{\Lambda}(t)$ must be replaced by $\tilde{\mathcal L}$ and $\vec\Xi(t)$ and the scalar product becomes the standard scalar product in the Euclidean space).
To this end, it is beneficial to choose $c=\lim_{t\rightarrow\infty}e_h(\chi(t))$ such that the second term in \eqref{eq:representation_of_G} vanishes asymptotically.
If such a matrix $\tilde{R}$ can be found and $G[\chi(t)]$ increases in first order, then  $G[\chi(t)]>0$ hold true for $t>0$ and the system dynamics is completely positive.

\subsection{Semi-definite programming}
\label{sec:semi-definite_programming}
The transformation introduced in \sref{sec:transformation_of_the_extended_dynamical_map}, or a consideration of more complex quantum systems and HEOMs with several levels, can quickly increase the dimensionality of the HEOM to a level at which the operator $\bm R$ can not be constructed analytically.
In such cases, however, the convexity of the problem permits the application of a highly efficient numerical approach.
To this end, let us parametrize $\bm R$ such that the normalization condition \ref{cond:normalization} (or \ref{condition:normalization_tp_equals_zero}) is satisfied and define $\bm{Q}\equiv\bm{\mathcal L}^\dagger \bullet{\bm R}+{\bm R}\bullet\bm{\mathcal L}$ and $\bm{B}=(v\mathbb{I}-\bm{Q}) \oplus (\bm{R}-\bm{S})$.
The minimization problem 
\be
v_m=\min_{v,\bm{R}}\left(v\left.\right| \bm{B}\ge 0\right)
\label{eq:sdp}
\ee
is a semi-definite program \cite{Vandenberghe1996}, which can be solved reliably and efficiently.
The constraint $\bm{B}\ge 0$ implies $\bm{R}-\bm{S} \ge 0$ (i.e.\ condition \ref{cond:consistency}) and $\bm{Q}\le v\mathbb{I}$.
If the minimum $v_m$ is non-positive, one has found $\bm{R}$ such that \ref{cond:monotonicity} holds true.
The condition $v_m \le 0$ thus verifies complete positivity.

\subsection{Example: non-Markovian dynamics due to a finite temperature bath}
\label{sec:example_non-markovian_bath_of_finite_temperature}
To gain a better intuition of how the transformation introduced in \sref{sec:transformation_of_the_extended_dynamical_map} affects the geometric properties of the problem, we consider a two-level system that is coupled to a bath of finite temperature.
In the Markovian case, the situation is described by a master equation in Lindblad form $\dot\varrho_1 = {\mathcal B}_+\varrho_1 + {\mathcal B}_- \varrho_1$ where 
\be
{\mathcal B}_\pm\varrho_1 = \gamma_\pm \left(\sigma_\pm\varrho_1\sigma_\mp- \left\lbrace \sigma_\mp\sigma_\pm,\varrho_1 \right\rbrace /2 \right)\ ,
\ee 
with rates $\gamma_\pm$.
In order to also permit non-Markovian dynamics, we introduce the traceless auxiliary operator $\varrho_1(t)$ and extend the Lindblad equation to the HEOM (see \ref{sec:extending_of_lindblad_equation} for a motivation of this equation)\\
\vbox{%
\numparts
\begin{equationarray}{rcr@{}rl}
\partial_t\varrho_1(t) &\,=\,& ({\mathcal B}_+ + {\mathcal B}_-) \,\varrho_1(t)&\,+\omega\,\varrho_2(t) \label{eq:bath_heom_a}\\
\partial_t\varrho_2(t) &\,=\,& \frac{\xi}{\gamma_+}({\mathcal B}_+ + {\mathcal B}_-)\,\varrho_1(t) &\,+\frac{\gamma_+ + \gamma_-}{2} ( {\mathcal D}_x + {\mathcal D}_y )\,\varrho_2(t)
\label{eq:bath_heom_b}
\end{equationarray}
\endnumparts
}
with $\gamma_p = (\gamma_+ + \gamma_-)/2$, $\gamma_m = (\gamma_+ - \gamma_-)/2$, and $\xi$ and $\omega$ being scalar parameters that characterize the additional levels in HEOM (the special cases $\omega=0$ and $\xi = 0$ give rise to Markovian dynamics).
With $\bm{\varrho} = \varrho_1\otimes\ket{1} + \varrho_2\otimes\ket{2}$ and the operator $\bm{\mathcal{L}}$ that is derived from the HEOM \eqref{eq:bath_heom_a}-\eqref{eq:bath_heom_b}, the latter can be expressed as $\partial_t\bm{\varrho}(t) = \bm{\mathcal L}\bullet\bm{\varrho}(t)$.
The equation of motion for the dynamical maps is obtained by replacing $\bm{\varrho}(t)$ with $\bm{\Lambda} = \Lambda_1\otimes\ket{1} + \Lambda_2\otimes\ket{2}$ that is initialized by $\Lambda_1(0)=\mathbb{I}$ and $\Lambda_2(0)=\mathbb{O}$.

The structure of $\bm{\mathcal L}$ and the initial condition give rise to a set of identities and restrictions which finally permit to fully represent $\bm{\Lambda}(t)$ by means of a four-dimensional vector $\vec\lambda(t)$, whose initial condition is $\vec\lambda(0)=[-\gamma_m/\gamma_p,0,1,0]$, and which evolves according to $\partial_t \vec\lambda(t) = l\, \vec\lambda(t)$ with 
\be
l=-
\left[
\begin{array}{cccc}
 2 \gamma_p & -\omega & 0 & 0 \\
 2 \xi  & 2 \gamma_p & 0 & 0 \\
 0 & 0 & \gamma_p & -\omega \\
 0 & 0 & \xi  & \gamma_p \\
\end{array}
\right]\ .
\ee

The dynamical map $\Lambda_1(t)$ in the Pauli basis is given as in \eqref{eq:Kraus} (with $\mu_i = \sigma_i$) and due to the structural properties of $\bm{\mathcal L}$ and $\bm{\Lambda}(0)$, the coefficient matrix is always of the form
\be
\label{eq:structure_chi_bath}
\chi(t)=
\left[
\begin{array}{cccc}
 \frac{2 \gamma_m \lambda_3(t) + \gamma_m - \gamma_p \lambda_1(t)}{4 \gamma_m} & 0 & 0 & \frac{\theta(t)}{4 \gamma_p} \\
 0 & \frac{\theta(t)}{4 \gamma_m} & \frac{-i \theta(t)}{4 \gamma_p} & 0 \\
 0 & \frac{i \theta(t)}{4 \gamma_p} & \frac{\theta(t)}{4 \gamma_m} & 0 \\
 \frac{\theta(t)}{4 \gamma_p} & 0 & 0 & \frac{-2 \gamma_m  \lambda_3(t) + \gamma_m - \gamma_p \lambda_1(t)}{4 \gamma_m} 
\end{array}
\right]\ ,
\ee
where $\theta(t) \equiv \gamma_m+\gamma_p\lambda_1(t)$.
In contrast to the previous examples, this structure of $\chi(t)$ does not permit conclusions about constantly vanishing eigenvalues such that the highest possible elementary symmetric polynomial $e_h = \det$ must be considered.
\Eref{eq:structure_chi_bath} does, however, give rise to the factorisation $e_h(\chi(t)) = \det(\chi(t)) = \alpha P_1^2(t)\, P_2(t)$ with a proportionality factor $\alpha = \gamma_+\gamma_-/[4(\gamma_-^2-\gamma_+^2)]^4$ and 
\numparts
\begin{eqnarray}
P_1(t) &= \gamma_p \lambda_1(t)  + \gamma_m \label{eq:P41} \\
P_2(t) &= c_1 -c_2^2\,\lambda_3^2(t) + c_3\,\lambda_1(t) + c_4\,\lambda_1^2(t)\ .
\end{eqnarray}
\endnumparts
Here, the constants are given by
$c_1 = \gamma_- \gamma_+ \gamma_m^2$,
$c_2=\gamma_-^2 - \gamma_+^2$,
$c_3 = \gamma_-^4 - \gamma_+^4$ and
$c_4 = \gamma_- \gamma_+ \gamma_p^2$.

\begin{figure}[t]
\centering
\label{fig:finiteTemp}
\includegraphics[height= 6cm]{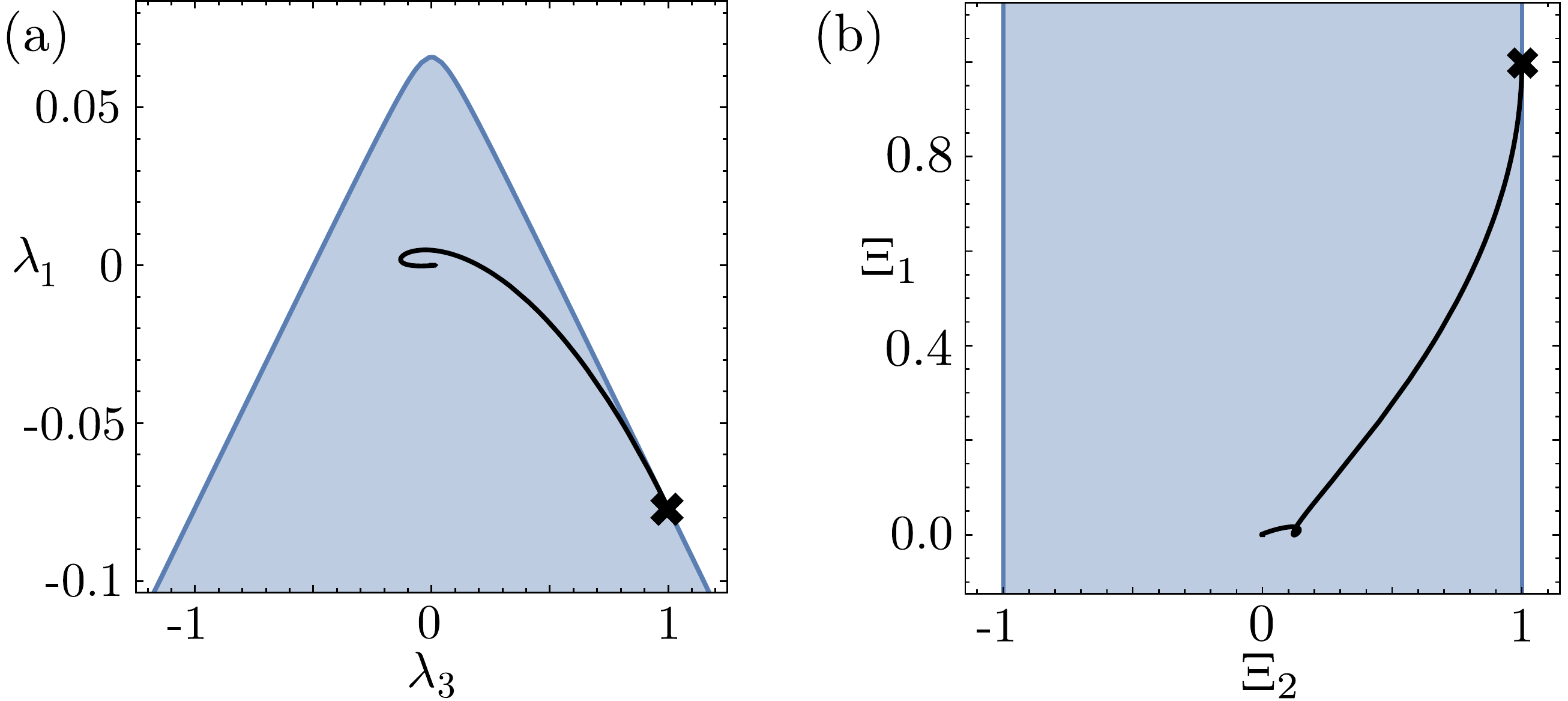}
\caption{The sets inside which $P_2 > 0$ is satisfied are depicted in blue before (a) and after (b) the non-linear transformation from $\vec\lambda(t)$ to $\vec\Xi(t)$. The black lines show the trajectories of $\vec\lambda(t)$ and $\vec\Xi(t)$ for the initial conditions that correspond to $\Lambda_1(0)=\mathbb{I}$ (black crosses). For the geometry depicted in (a), it is impossible to find a matrix $r^{(2)}$ such that the equipotential line of points $\vec\xi$ with $\vec\xi^\dagger\,r^{(2)}\,\vec\xi=\textrm{\it const}$ contains $\vec\lambda(0)$ and lies completely inside ${\cal P}_2$ (the set with $P_2>0$). After the non-linear transformation a more benign geometry has been obtained such that an according matrix can be found.}
\end{figure}
Considering the case $\gamma_+,\gamma_- > 0$, the proportionality factor $\alpha$ is positive such that it is sufficient to show strict positivity of $P^2_1(t)$ and $P_2(t)$ separately in order to prove the positivity of $e_h(\chi(t))$. 
After verifying strict positivity of $\chi(t)$ for short times, this ensures CP dynamics for $t \ge 0$.

The factor $P^2_1(t)$ is always non-negative but to show strict positivity for $t>0$, we reformulate the problem such that $\vec\lambda^\dagger(0)\,s^{(1)}\,\vec\lambda(0)-\vec\lambda^\dagger(t)\,s^{(1)}\,\vec\lambda(t)>0$ with a symmetric matrix $s^{(1)}$ is sufficient for $P_1^2(\vec\lambda(t))>0$ and strive for the construction of a matrix $r^{(1)}$ that satisfies $l^\dagger\,r^{(1)} + r^{(1)}\,l \le 0$,  $\vec\lambda^\dagger(0)\,(r^{(1)}-s^{(1)})\,\vec\lambda(0)=0$, and $r^{(1)}-s^{(1)} \ge 0$ (see \ref{cond:monotonicity}, \ref{condition:normalization_tp_equals_zero}, and \ref{cond:consistency}).
The procedure is very similar to the previous examples for why we abstain from a detailed discussion and anticipate that $P_1(t)$ never returns to zero when $-2 \gamma_p^2\leq \omega\xi$ holds true.

Showing the second condition $P_2(\vec\lambda(t)) = \vec\lambda^\dagger(0)\,s^{(2)}\,\vec\lambda(0) -  \vec\lambda^\dagger(t)\,s^{(2)}\,\vec\lambda(t)> 0$ is more involved, which can be understood best in geometric terms.
A cross-section (for $\lambda_2=\lambda_4=0$) of the set ${\cal P}_2$ of vectors $\vec\lambda$ that satisfy $P_2(\vec\lambda) > 0$ is depicted in \fref{fig:finiteTemp}(a) and the black cross corresponds to the initial condition $\vec\lambda(0)$.
As one can see, it is not possible to place an ellipsoidal equipotential line $\vec\lambda^\dagger\,r^{(2)}\,\vec\lambda = \textrm{\it const}$ (that would confine the set ${\cal V}_2$) inside ${\cal P}_2$ such that it contains the initial condition $\vec\lambda(0)$.
In algebraic terms, the two conditions \ref{condition:normalization_tp_equals_zero} and \ref{cond:consistency} can not be satisfied simultaneously for geometric reasons.

We can, however, transform the geometry of ${\cal P}_2$ in order to mitigate this problem.
To this end, we define the first component of the new vector $\vec\Xi(t)$ to be given by $\Xi_1(t) = c - P_2(t)$ with $c=\lim_{t\rightarrow\infty}(P_2(t))>0$ such that $G \equiv \vec\Xi^\dagger(0)\,\tilde{S}\,\vec\Xi(0) - \vec\Xi^\dagger(t)\,\tilde{S}\,\vec\Xi(t) > 0$ implies $P_2(t)>0$, when $\tilde{S}=\textrm{diag}[1,0,\ldots,0]$.
The other elements $\Xi_j(t)=f_j(\lambda_k(t),\lambda_k(t)\,\lambda_l(t))$ with $j \ge 1$ are polynomials in $\lambda_k(t)$ of degree one or two and are iteratively chosen such that $\vec\Xi(t)$ satisfies the differential equation $\partial_t\vec{\Xi}(t) = \tilde{\mathcal L}\,\vec\Xi(t)$ with an according generator $\tilde{\mathcal{L}}$.

\Fref{fig:finiteTemp}(b) depicts the convex set of vectors $\vec\Xi$ for which $G > 0$ holds true.
Because of its benign geometry, we can find a matrix $\tilde R$ such that the equipotential line $\vec\Xi^\dagger\,\tilde R\,\vec\Xi = \textrm{\it const}$ contains the initial condition $\vec\Xi(0)$ and lies completely within ${\cal P}_2$, i.e.\ $\vec\Xi^\dagger(0)\,(\tilde R - \tilde S)\,\vec\Xi(0)=0$ and $\tilde R - \tilde S \ge 0$ hold true.
When also the monotonicity condition $\tilde{\mathcal L}^\dagger\,\tilde R + \tilde R\,\tilde{\mathcal L} \le 0$ is satisfied, then this proves $P_2(t) > 0$ for all times.

Due to the rather high dimensionality of $\vec\Xi(t)$, such a function can not be found analytically but by means of a semi-definite program a very efficient numeric construction has been carried out.

\section{Quantifying the violation of complete positivity}
\label{sec:violation_of_cp}
In many cases, an exact description of the dynamics of a physical system is only obtained for infinitely deep hierarchy equations.
A truncation of the HEOM is then often employed in order to obtain approximate system dynamics.
Such a truncation, however, can give rise to a slight violation of complete positivity of the induced dynamical system map $\Lambda_1(t)$.
The latter typically occurs for short times, at which the coefficient matrix $\chi(t)$ has very small eigenvalues, and vanishes for all times greater than a critical time $t_p$.
During this initial time interval $0 \le t \le t_p$ complete positivity is violated and even if the violation is numerically negligible, the proposed framework will not be able to verify CP dynamics.

In such cases, it is desirable to estimate the degree of violation of complete positivity. 
This degree is for instance an indicator of whether the approximation is good enough as strong violation of complete positivity suggests unphysical dynamics and thus an insufficient approximation.
We will propose two different ways of quantifying this violation and demonstrate their applicability by means of an important example.

\subsection{Complete positivity after an initial violation time}
\label{sec:complete_positivity_after_an_initial_violation_time}
At the initial time $t=0$, when the system map $\Lambda_1(0)$ is equal to the identity, the coefficient matrix $\chi(0)$ has just one non-vanishing eigenvalue.
A truncation (and thus approximation) of the HEOM therefore often causes an eigenvalue of $\chi(t)$ to become negative for short times.
The identification of a critical time $t_p$ at which the dynamics becomes CP again helps to estimate whether the violation is indeed just a short-time phenomenon, or a structural problem that eventually impedes a physical interpretation of the time-evolution.
A good guess for $t_p$ can be obtained from a numeric propagation of $\chi(t)$ for short times and once a candidate for $t_p$ (and the according extended map $\bm{\Lambda}(t_p)$) has been identified, the proposed method in \sref{sec:the_algebraic_framework} for proving complete positivity can be employed to show that the dynamics is CP for all $t > t_p$.

\begin{figure}[t]
\centering
\includegraphics[height= 6cm]{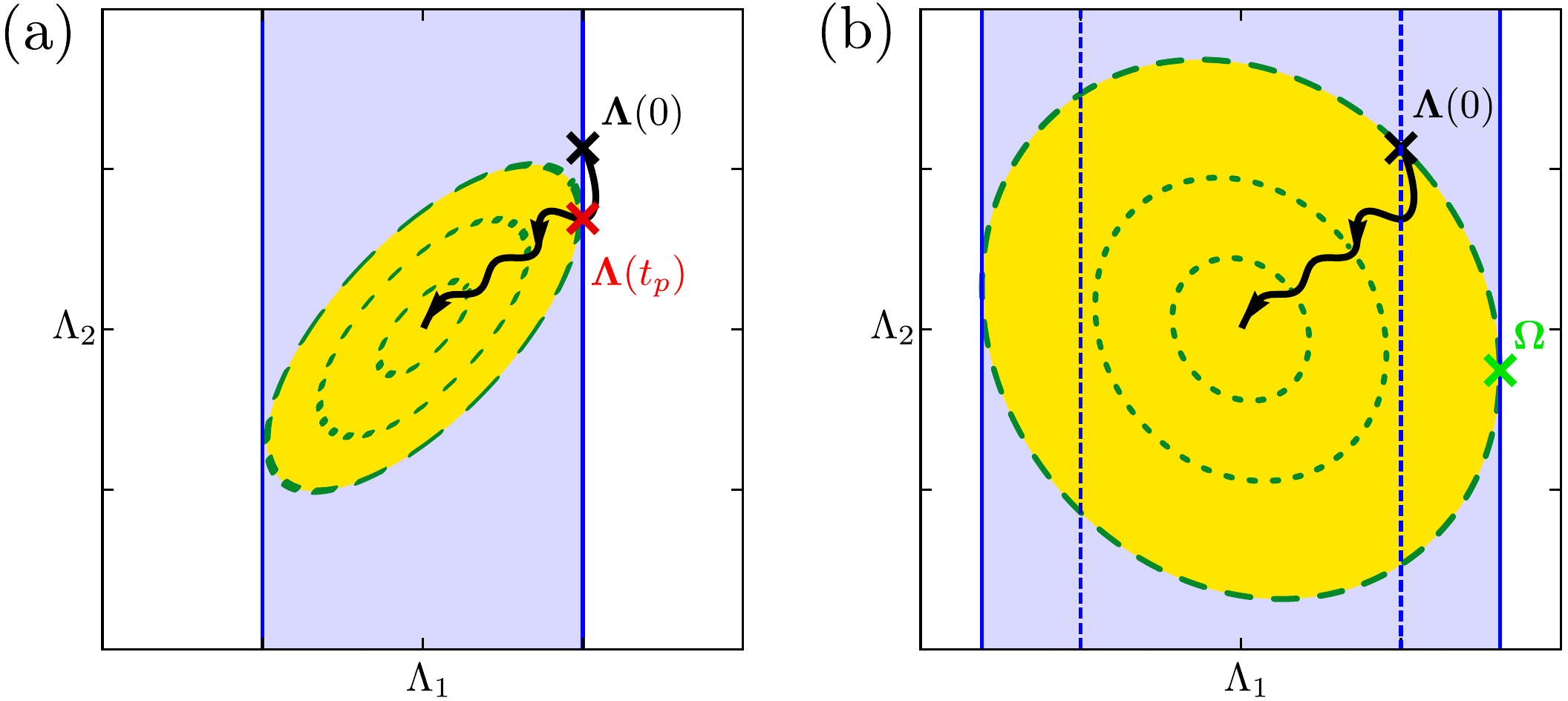}
\caption{The figure shows schematically how a time-evolution violates complete positivity for short times before the dynamics becomes CP again. The induced dynamics that is initialized by $\bm{\Lambda}(0)$ (black cross) is represented by the black trajectory. The valid region $\cal P$ of extended maps that induce CP dynamics is depicted in blue and $\bm{\Lambda}(t)$ leaves this region for short times. In (a), the initial condition is thus replaced by $\bm{\Lambda}(t_p)$ (red cross). The time interval  $0 \le t \le t_p$ is disregarded and the proposed procedure is employed to determine the set $\cal V$, that contains $\bm{\Lambda}(t)$ for $t>t_p$, such that $\cal V$ lies inside $\cal P$ and  complete positivity is shown for $t>t_p$. In (b), the condition $\chi(t) > 0$, that defines $\cal P$, is replaced by $\chi(t) >-\delta_{min}\mathbb{I}$ which enlarges the set $\cal P$ of states that are considered valid. The tangential point $\bm{\Omega}$ of $\cal P$ and $\cal V$ becomes object to an optimization, which increases the chance to find $\cal V$ such that ${\cal V} \subseteq {\cal P}$.
In that case, $-\delta_{min}$ is a lower bound for the eigenvalues of $\chi(t)$.}
\label{fig:cp_violation}
\end{figure}

The situation is depicted in \fref{fig:cp_violation}(a), where it is schematically visualized how the extended dynamical map leaves the valid region $\cal P$ for short times in order to re-enter at time $t_p > 0$.

\subsection{Lower bound on the eigenvalues of $\chi$}
\label{sec:lower_bound_for_the_eigenvalues_of_chi}
An alternative way of quantifying the violation of complete positivity of the system map $\Lambda_1(t)$ is based on estimating a lower bound $-\delta_{min}$ on the eigenvalues of the matrix $\chi(t)$.
Rather than requiring the positivity of the highest non-trivial elementary symmetric polynomial $e_h(\chi(t))$, we strive to prove $e_h(\chi(t) + \delta_{min}\mathbb{I})>0$, which permits negative eigenvalues of $\chi(t)$ as long as they remain greater than the new parameter $-\delta_{min}$.

Algebraically this means that one needs to find an extended map $\bm\Omega=\sum_i\Omega_i\otimes\ket{i}$ that is such that \eqref{eq:representation_det_chi} is modified to
\be
e_h(\chi(t) + \delta_{min}\mathbb{I})=\langle \bm{\Omega}, \bm{S}\bullet\bm{\Omega}\rangle-\langle\bm{\Lambda}(t), \bm{S}\bullet\bm{\Lambda}(t)\rangle\ .
\label{eq:definition_extended_determinant}
\ee
Furthermore, the normalization condition~\ref{cond:normalization} is then replaced by
\begin{enumerate}[label=(\roman*'')]
\setcounter{enumi}{1}
\item  $\langle\bm{\Omega} ,(\bm{R}-\bm{S})\bullet\bm{\Omega}\rangle = 0$
\label{condition:normalization2} 
\end{enumerate}
and the set of conditions on $\bm R$ is extended by
\begin{enumerate}[label=(\roman*)]
\setcounter{enumi}{3}
\item $\langle\bm{\Omega}, \bm{R}\bullet\bm{\Omega}\rangle > \langle  \bm{\Lambda}(0), \bm{R}\bullet\bm{\Lambda}(0)\rangle$\ .
\label{cond:init_in_v}
\end{enumerate}
Whenever $\bm R$ is found such that \ref{cond:monotonicity}, \ref{condition:normalization2}, \ref{cond:consistency}, and \ref{cond:init_in_v} are satisfied, this proves strict positivity of $e_h(\chi(t + \delta_{min}))$ for all times $t \ge 0$.

To prove this statement, \ref{condition:normalization2} is applied to \eqref{eq:definition_extended_determinant}, which yields
\be
e_h(\chi(t) + \delta_{min}\mathbb{I}) = \langle\bm{\Omega}, \bm{R}\bullet\bm{\Omega}\rangle -  \langle\bm{\Lambda}(t),\bm{S}\bullet\bm{\Lambda}(t)\rangle\ .
\ee
By means of \ref{cond:consistency}, this can be further bound to
\be
e_h(\chi(t) + \delta_{min}\mathbb{I}) \ge \langle \bm{\Omega},\bm{R}\bullet\bm{\Omega}\rangle -  \langle\bm{\Lambda}(t),\bm{R}\bullet\bm{\Lambda}(t)\rangle\ ,
\label{eq:definition_extended_determinant_step_2}
\ee
which, according to \ref{cond:init_in_v}, is strictly positive at the time $t = 0$.
For all times $t > 0$, the right-hand side of \eqref{eq:definition_extended_determinant_step_2} increases monotonically due to \ref{cond:monotonicity}, such that $e_h(\chi(t) + \delta_{min}\mathbb{I}) > 0$ and all eigenvalues of $\chi(t)$ are greater than $-\delta_{min}$ for all times.

In geometric terms, this modification enlarges the region $\cal P$ that is considered valid as depicted in \fref{fig:cp_violation}(b).
The points $\bm\Gamma$ with ${\cal F}(\bm{\Gamma}) \le {\cal F}(\bm{\Omega})$ (where $\cal F$ is defined as in \sref{sec:geometric_interpretation}) define the region $\cal V$ that will contain the solution $\bm{\Lambda}(t)$ for all times and $\bm{\Omega}$ is the tangential point of $\cal V$ and the (enlarged) set $\cal P$.
It is important to point out that $\bm{\Omega}$ is not uniquely defined and becomes itself object to an optimization.
This often permits to find an operator $\bm R$ that satisfies \ref{cond:monotonicity}, \ref{condition:normalization2}, \ref{cond:consistency}, and \ref{cond:init_in_v} and proves $-\delta_{min}$ to be a lower bound on the eigenvalues of $\chi(t)$.

\subsection{Example: The spin-Boson model}
\label{sec:example_the_spin-boson_model}
One of the most prominent examples of a microscopically derived HEOM describes the dynamics of a two-level system that is coupled to a bath of harmonic oscillators.
For a high temperature case ($\beta\gamma \ll 1$) with weak system-bath coupling ($\Delta^2/\gamma^2 \ll \omega$), the HEOM is expressed as \cite{Tanimura2014,Tanimura2015,Tanimura2006}\\
\vbox{
\numparts
\begin{equationarray}{rcr@{}c@{}lr@{}c@{}l@{}}
\dot{\varrho}_1 &=&-i\frac{\omega}{2}[\sigma_z,&\varrho_1&]&-i\Delta[\sigma_x,&\varrho_2&] \label{eq:hierarchyExampleTanimura_a}\\
\dot{\varrho}_2 &=&-i\Delta[\sigma_x,&\varrho_1&]-\frac{\Delta\beta\gamma}{2}\{\sigma_x,\varrho_1\}&-i\frac{\omega}{2}[\sigma_z,&\varrho_2&]-\gamma\varrho_2\ .\label{eq:hierarchyExampleTanimura_b}
\end{equationarray}
\endnumparts
}
Here, $\varrho_1(t)$ represents the state of a system with a resonance frequency $\omega$, whereas $[\cdot,\cdot]$ and $\lbrace\cdot,\cdot\rbrace$ denote the commutator and the anti-commutator, respectively.
As discussed in \cite{Tanimura2014,Tanimura2015,Tanimura2006}, $\Delta$ characterizes the system-environment coupling, $\beta$ corresponds to the inverse bath temperature, and $\gamma$ is a damping constant.

Again \eref{eq:hierarchyExampleTanimura_a}/\eqref{eq:hierarchyExampleTanimura_b} can be written as $\partial_t\bm{\varrho}(t) = \bm{\mathcal L}\bullet\bm{\varrho}(t)$, where $\bm{\varrho} = \varrho_1\otimes\ket{1} + \varrho_2\otimes\ket{2}$ and the generator $\bm{\mathcal L}$ has been deduced form the HEOM.
In order to obtain the equation of motion for the extended dynamical map $\bm{\Lambda}(t)$, one replaces $\bm{\varrho}(t)$ by $\bm{\Lambda}(t)=\Lambda_1\otimes\ket{1} + \Lambda_2\otimes\ket{2}$.
Expanding the eigenvalues $\epsilon_i(t)$ of the matrix $\chi(t)$ that characterizes the dynamical map $\Lambda_1(t)$ for the system state in time, one finds that one of these eigenvalues is given by $\epsilon_1(\chi(t)) = -\frac{1}{144}\left(\beta^{2}\gamma^{2}\Delta^{2}\omega^{2}\right)t^{4}+O(t^{5})$.
For non-vanishing values of $\beta$, $\gamma$, $\Delta$ and $\omega$, this eigenvalue becomes negative for short times, which results in a violation of complete positivity of $\Lambda_1(t)$. 
Sufficient conditions for complete positivity can hence not be found for this important example.
It is, however, possible to estimate the degree of violation of complete positivity: 
A numerical integration yields that the violation is typically not very strong and limited to rather short times, both of which we prove in the following.

\subsubsection{Complete positivity for $t>t_p$}
\label{sec:complete_positivity_for_t_g_t_p}
By propagating the extended dynamical map over a short period of time, it is possible to determine a time $t_p$ at which the coefficient matrix $\chi(t_p)$ is positive definite. 
Based on the procedure laid out in \sref{sec:complete_positivity_after_an_initial_violation_time}, we want to prove that complete positivity is given for all subsequent times.

Due to the structure of $\bm{\mathcal L}$ and $\bm{\Lambda}(0)$, it is possible to represent the extended dynamical map $\bm{\Lambda}(t)$ by six degrees of freedom that form the six-dimensional real-valued vector $\vec\lambda(t)$, which is initialized by $\vec\lambda(0)=[1,0,1,0,0,1]$. The latter evolves according to the differential equation $\partial_t \vec\lambda(t) = l\,\vec\lambda(t)$ with
\be
\label{eq:reducedDifferentialOperatorTanimura}
l = \left[
\begin{array}{cccccc}
 0 & \omega & 0 & 0 & 0 & 0 \\
 0 & 0 & -\omega & 0 & 0 & 0 \\
 \gamma  & \frac{4 \Delta ^2}{\omega}+\omega & -\gamma  & 0 & 0 & 0 \\
 0 & 0 & 0 & -\gamma  & -\omega & 0 \\
 0 & 0 & 0 & \omega & -\gamma  & -2 \Delta  \\
 0 & 0 & 0 & 0 & 2 \Delta  & 0 \\
\end{array}
\right].
\ee
The simplification of $\bm{\Lambda}(t)$ yields a set of identities for the dynamical map $\Lambda_1(t)$, which gives rise to in a certain structure of $\chi(t)$ given by
\be
\label{eq:coefficientMatrixStructureTanimura}
\chi(t)=
\left[
\begin{array}{cccc}
\chi_{11}(t) & 0 & 0 & \chi_{14}(t) \\
 0 & \chi_{22}(t) & \chi_{23}(t) & 0 \\
 0 & \chi_{23}^*(t) & \chi_{33}(t) & 0 \\
 \chi^*_{14}(t) & 0 & 0 & \chi_{44}(t) \\
\end{array}
\right]\ ,
\ee
where all matrix elements are linear functions in $\vec\lambda(t)$.
The structure of $\chi(t)$ does not permit conclusions concerning the existence of constantly vanishing eigenvalues such that the determinant is considered to be the highest non-trivial elementary symmetric polynomial.
It is evaluated to $e_h(\chi(t)) = \det(\chi(t)) \propto P_1(t) P_2(t)$ with a positive proportionality factor and
\numparts
\begin{eqnarray}
\fl
P_1 &=  -4 \beta ^2 \Delta ^2 \lambda_4^2-4 \beta^2 \Delta  \omega (\lambda_6-1) \lambda_4
-(\lambda_6-2) \lambda_6 \left(\beta ^2 \omega^2-4\right)-4 (\lambda_1-\lambda_3)^2+b\label{eq:finitePropagationHighestPolynomial_a}\\
\fl P_2 &= -4 \left(4 \lambda_2^2+(\lambda_1+\lambda_3)^2-(\lambda_3+1)^2\right)-(2 \beta  \Delta  \lambda_4+\beta  \omega (\lambda_6-1))^2\ , \label{eq:finitePropagationHighestPolynomial_b}
\end{eqnarray}
\endnumparts
where $b = -\beta ^2 \omega^2+4$.
For the considered initial condition $\vec\lambda(0)$, both factors vanish initially and by algebraic means it can be shown that $P_1(t)=P_2(t)$ holds indeed true for all times $t \ge 0$.
It is thus sufficient to keep track of $P_1(t)$ only, and to show that it never returns to zero, in order to prove $e_h(\chi(t)) > 0$ and thus complete positivity of $\Lambda_1(t)$ for $t \ge t_p$.

For geometrical reasons that are similar to those discussed in \sref{sec:example_non-markovian_bath_of_finite_temperature}, the transformation that has been introduced in \sref{sec:transformation_of_the_extended_dynamical_map} is required in order to render a situation in which our procedure is applicable.
To this end, a vector $\vec\Xi(t)$ is constructed based on $\vec\lambda(t)$ that is such that $G = \vec\Xi^\dagger(0)\,\tilde{S}\,\vec\Xi(0) - \vec\Xi^\dagger(t)\,\tilde{S}\,\vec\Xi(t) > 0$ (with the matrix $\tilde{S}=\textrm{diag}[1,0,\ldots,0]$) implies $P_1(t)>0$ for $t > t_p$.
The strict positivity of $G(t)$ is proven when a matrix $\tilde{R}$ has been found that satisfies $\tilde{\mathcal L}^\dagger\,\tilde{R}+\tilde{R}\,\tilde{\mathcal L} \le 0$, $\vec{\Xi}^\dagger(t_p)\,(\tilde{R}-\tilde{S})\,\vec\Xi(t_p) = 0$, and $\tilde{R}-\tilde{S} \ge 0$, which can be carried out by means of a semi-definite program.

\begin{figure}
\centering
\includegraphics[scale=.35]{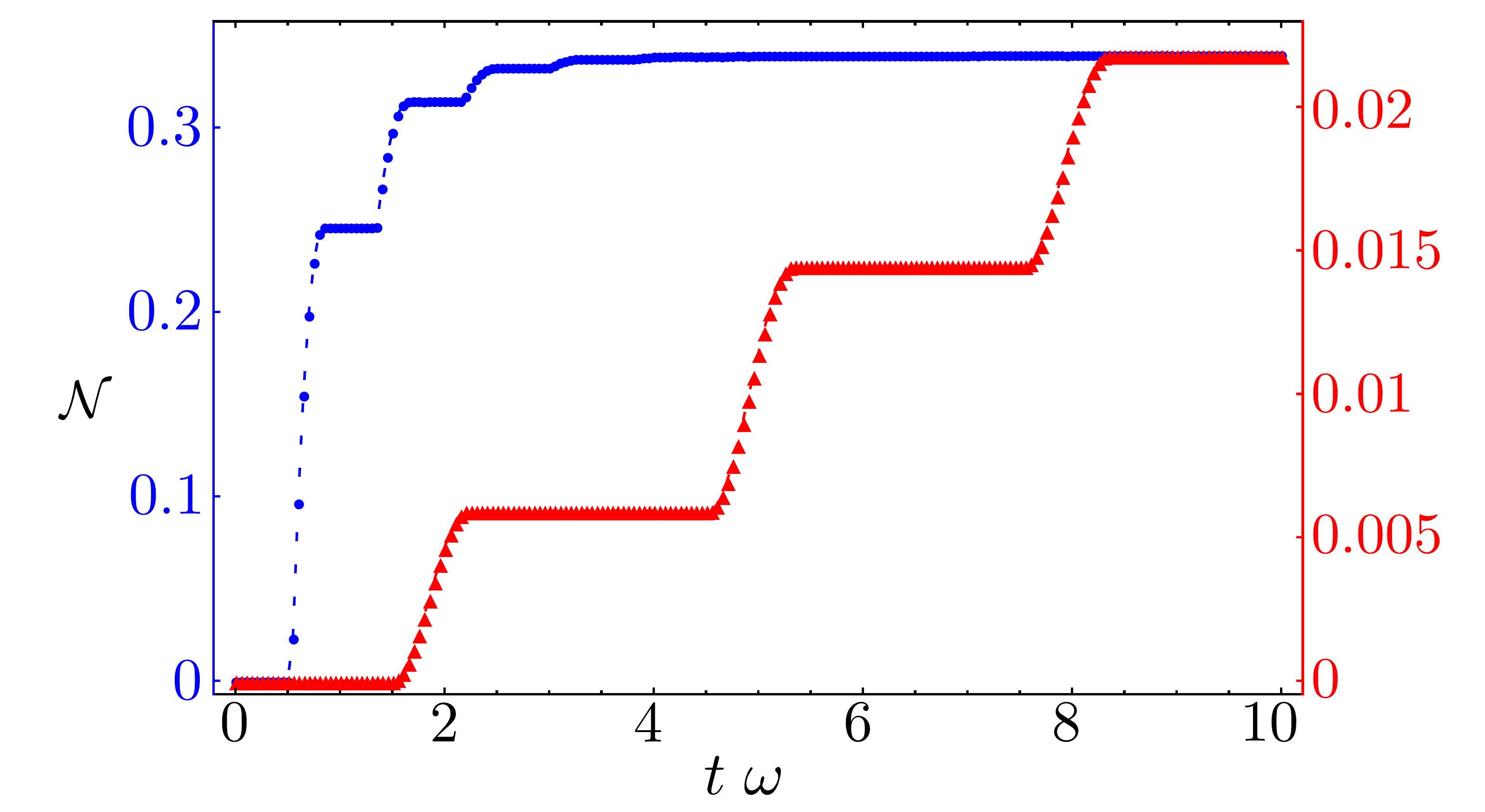}
\caption{A measure of non-Markovianity \cite{Breuer2009a} ${\cal N}(\Lambda_1(t))$ is depicted as a function of time for two different numerical instances of the HEOM in \eqref{eq:hierarchyExampleTanimura_a} and \eqref{eq:hierarchyExampleTanimura_b}. The blue dots represent the non-Markovianity for a HEOM determined by $\gamma=3\,\omega$, $\Delta=2\,\omega$ and $\beta=8/10\,\omega^{-1}$, which converges to a non-Markovianity of ${\cal N}(\Lambda_1(t\rightarrow\infty)) = 0.34$. Our framework proves the dynamics to be CP for all times greater then $t_p \simeq 0.6/\omega$. The red triangles relate to the case of higher temperatures ($\gamma=1\,\omega$, $\Delta=2/10\,\omega$ and $\beta=2/10\,\omega^{-1}$), for which the total non-Markovianity is measured to ${\cal N}(\Lambda_1(t\rightarrow\infty))=0.08$. As we show, none of the eigenvalues of the matrix $\chi(t)$ ever falls below $-\delta_{min}=-10^{-2}$ such that CP-violation is moderate.}
\label{fig:traceDistancesExamples}
\end{figure}

As an example, we consider the case $\gamma = 3\,\omega$, $\Delta = 2\,\omega$ and $\beta = 8/10\,\omega^{-1}$ depicted in \fref{fig:traceDistancesExamples}. 
This case features a clear signature of non-Markovianity, which is measured \cite{Breuer2009a} to ${\cal N}(\Lambda_1(t\rightarrow\infty)) = 0.34$.
Complete positivity is proven after the critical time $t_p\simeq 0.6/\omega$.
It is noteworthy that the non-Markovian revivals arise after the critical time $t_p$.

\subsubsection{Lower bound on the eigenvalues of $\chi(t)$}
The second possibility to quantify the violation of complete positivity is to estimate a lower bound $-\delta_{min}$ on the eigenvalues of the matrix $\chi(t)$ as discussed in \sref{sec:lower_bound_for_the_eigenvalues_of_chi}. 
To this end, we make a guess for $\delta_{min} > 0$ and prove $e_h(\chi(t) + \delta_{min}\mathbb{I}) = \det(\chi(t) + \delta_{min}\mathbb{I}) >0$ for $t \ge 0$.
The time-evolution of $\bm{\Lambda}(t)$ is not affected by this modification of $e_h$ for why the generator $l$ given in \eqref{eq:reducedDifferentialOperatorTanimura} and the structure of the matrix $\chi(t)$ in \eqref{eq:coefficientMatrixStructureTanimura} still apply.

The modified polynomial $e_h$ can again be factorized to $e_h(\chi(t) + \delta_{min}\mathbb{I}) \propto P'_1(t) P'_2(t)$ with a positive proportionality constant and\\
\vbox{
\numparts
\begin{eqnarray}
P'_1(t) = & P_1(t) + 32\delta_{min}(1 + 2\delta_{min} - \lambda_6(t)) \label{eq:FiniteViolationHighestPolynomial_a}\\
P'_2(t) = & P_2(t) + 32\delta_{min}(1 + 2\delta_{min} + \lambda_6(t))\ , \label{eq:FiniteViolationHighestPolynomial_b}
\end{eqnarray}
\endnumparts
}
where $P_{1,2}(t)$ are given in \eqref{eq:finitePropagationHighestPolynomial_a} and \eqref{eq:finitePropagationHighestPolynomial_b}. 
In contrast to the previous case, the two factors $P'_{1,2}(t)$ differ from each other for why they must be examined individually.

The procedure to prove $P'_{1,2}(t) > 0$ is similar to the previous case in \sref{sec:complete_positivity_for_t_g_t_p}. 
Again, transformations are applied that give rise to new vector representations $\vec\Xi^{(1)}(t)$ and $\vec\Xi^{(2)}(t)$ for the extended dynamical map and the according generators $\tilde{\mathcal L}^{(i)}$ (with $i=1,2$) which govern the dynamics.
Matrices $\tilde S^{(i)}$ and functions $G_i(t) \equiv [\vec\Omega^{(i)}]^\dagger\,\tilde{S}^{(i)}\,\vec\Omega^{(i)}
 - [\vec\Xi^{(i)}(t)]^\dagger\,\tilde{S}^{(i)}\,\vec\Xi^{(i)}(t)$  are defined such that $G_i(t)>0$ implies $P'_i(t) > 0$ for all times.
The identification of matrices $\tilde{R}^{(i)}$ that satisfy $[\tilde{\mathcal L}^{(i)}]^\dagger\,\tilde{R}^{(i)} + \tilde{R}^{(i)}\,\tilde{\mathcal L}^{(i)} \le 0$ (condition \ref{cond:monotonicity}), are normalized as $[\vec\Omega^{(i)}]^\dagger\,(\tilde{R}^{(i)}-\tilde{S}^{(i)})\,\vec\Omega^{(i)}=0$ (condition \ref{condition:normalization2}), satisfy the operator inequalities  $\tilde{R}^{(i)}-\tilde{S}^{(i)} \ge 0$ (condition \ref{cond:consistency}), and the additional conditions $[\vec\Xi^{(i)}(0)]^\dagger\,\tilde{R}^{(i)}\,\vec\Xi^{(i)}(0) < [\vec\Omega^{(i)}]^\dagger\,\tilde{R}^{(i)}\,\vec\Omega^{(i)}$ (condition \ref{cond:init_in_v}) do then prove $G_i(t)$ to be strictly positive, which implies $e_h(\chi(t) + \delta_{min})>0$ such that $-\delta_{min}$ is a lower bound on the eigenvalues of $\chi(t)$.

As mentioned before, the vectors $\vec\Omega^{(i)}$ are not uniquely defined and become objects to an optimization. 
Good candidates are determined with a gradient-based method and by means of a semi-definite program we show that $-\delta_{min}=-10^{-2}$ is a lower bound on the smallest eigenvalue of $\chi(t)$ for the test case of $\gamma = 1\,\omega$, $\Delta = 2/10\,\omega $ and $\beta = 2/10\,\omega^{-1}$.
This HEOM induces a dynamical process with a non-Markovianity of ${\cal N}(\Lambda_1(t \rightarrow \infty)) = 0.08$, whose time evolution is depicted in \fref{fig:traceDistancesExamples}.

\section{Conclusion}
\label{sec:conclusion}
We have demonstrated how complete positivity can be incorporated into the framework of HEOMs. This enables an abstraction of the equations of motion
from a derivation based on microscopic models towards a more phenomenological perspective that opens a new approach to non-Markovianity.
A construction of elementary models of open system dynamics in which physical quantities like amplitudes and frequencies of revivals of quantum coherence have a clear root in the underlying equations of motion -- just like decay constants are rooted in Markovian master equations -- will ultimately ease investigations of open system dynamics in which system size or large statistical sampling \cite{Walschaers2015,Scholak2011, Mostarda2013, Witt2013} requires efficient phenomenological models.
Although a theorem that describes CP dynamics as beautifully as the Lindblad theorem in the Markovian case
is currently far out of reach for non-Markovian systems, the presented approach might be taken as an initial step towards such a theory.

\ack
We are indebted to Heinz-Peter Breuer, Akihito Ishizaki and Robert Alicki for stimulating discussions.
Financial support by the ERC within the project ODYCQUENT is gratefully acknowledged.

  \begin{appendix}
  \section{Phenomenological constructions of HEOMs}
  In \sref{sec:introductory_examples}, \sref{sec:example_non-markovian_bath_of_finite_temperature}, and \sref{sec:example_the_spin-boson_model} specific examples of HEOMs have been examined with respect to complete positivity.
  Whereas the HEOM in \eqref{eq:hierarchyExampleTanimura_a}/\eqref{eq:hierarchyExampleTanimura_b}, which has been discussed in \sref{sec:example_the_spin-boson_model}, is a well-known HEOM for the spin-boson model and has first been derived in \cite{Tanimura1989a} based on microscopic assumptions, the other examples have been obtained from phenomenological considerations rather than from microscopic system-environment models.
  To this end, new techniques have been conceived that will be presented in \ref{sec:targeted_solution} and \ref{sec:extension_of_lindblad} before the HEOMs that have been investigated in the main article will be derived explicitly.
  
  \subsection{Finding HEOMs for targeted solutions}
  \label{sec:targeted_solution}
  A convenient and constructive way to obtain a well-behaved HEOM is to first specify the targeted system dynamics and then tailor the HEOM according to this solution.
  To this end, one first defines a dynamical map $\Lambda_1(t)$ that propagates the system state $\varrho_1(t) = \Lambda_1(t)\,\hat\varrho$ for all times $t \ge 0$ and all initial states $\varrho_1(0) = \hat\varrho$.
The HEOM will then be constructed such that $\Lambda_1(t)\,\hat\varrho$ is the first element of the extended state $\bm{\varrho}(t)=\sum_{i=1}^n \varrho_i(t)\otimes\ket{i}$ that solves the HEOM for the initial condition $\varrho_1(0)=\hat\varrho$ and $\varrho_i(0)=\mathbb{O}$ for all $i \ge 2$.

The HEOMs that we target for are of triangular form, {\it i.e.}\ their generators satisfy
\be
{\mathcal L}_{ij} = \omega \mathbb{I}
\hspace{5px} \textrm{ for } \hspace{5px}
j=i+1 \hspace{15px} \textrm{ and } \hspace*{15px} {\mathcal L}_{ij} = \mathbb{O}
\hspace*{5px} \textrm{for} \hspace*{5px} j \ge i + 2
\label{eq:hierarchy_operator}
\ee
with $1/\omega$ being a unit of time and $\mathbb{I}$ and $\mathbb{O}$ being the identity map and the null map, respectively.
With this form of $\bm{\mathcal L}$, the level $k$ of the HEOM can affect the system dynamics earliest in the $k^{th}$ time-step.

We will now successively deduce the yet unknown operators ${\mathcal L}_{ij}$, with $j \le i$ by imposing the condition that each operator $\varrho_k(t)$ must vanish in the first $k-2$ time-steps:
Whereas the system state $\varrho_1(t)$ does not have to vanish at all, the first auxiliary operator $\varrho_2(t)$ shall vanish at the time $t=0$, the second auxiliary operator $\varrho_3(t)$ shall satisfy $\varrho_3(0)=\mathbb{O}$ as well as $\partial_t\varrho_3(t)|_{t=0}=\mathbb{O}$, and this series of conditions will be continued for all $\varrho_k(t)$.

Let us demonstrate the procedure more explicitly:
After having defined a dynamical map $\Lambda_1(t)$ for the system state, we define 
\be
\tilde\varrho_2({\mathcal L_{11}},\hat\varrho,t) \equiv\left( \dot\Lambda_1(t)\hat\varrho - {\mathcal L}_{11}\Lambda_1(t)\,\hat\varrho\right)/\omega\ ,
\label{eq:definitition_of_tilde_rho_1}
\ee
as a function of time, the initial system state $\hat\varrho$, and the yet unknown operator ${\mathcal L}_{11}$.
We then require $\tilde\varrho_2({\mathcal L_{11}},\hat\varrho,t)$ to vanish at the initial time $t=0$ independently of the initial system state $\hat\varrho$, which can be expressed by
\be
\frac{\partial}{\partial\hat\varrho_{ij}} \tilde\varrho_2({\mathcal L}_{11},\hat\varrho,0) = 0\hspace*{1cm}\forall i,j\ .
\label{eq:determination_of_l_00}
\ee
Solving this linear equation with respect to ${\mathcal L}_{11}$ determines the first level of the HEOM as well as the first auxiliary operator, which can now be obtained from $\dot\varrho_1(t)={\mathcal L}_{11}\,\varrho_1(t) + \omega\,\varrho_2(t)$ and reads
\be
\varrho_2(t) \equiv (\dot\varrho_1(t) - {\mathcal L}_{11}\,\varrho_1(t))/\omega \ .
\label{eq:definitition_of_rho_0}
\ee

In order to obtain the operators ${\mathcal L}_{k1}$ to ${\mathcal L}_{kk}$ for each higher level $k$ of the HEOM, we iteratively reapply this procedure.
That is, we first define a function
\be
\tilde\varrho_{k+1}({\mathcal L}_{k1},\ldots,{\mathcal L}_{kk},\hat\varrho,t) \equiv \left(\dot\varrho_k(t) - \sum_{j=1}^{k}{\mathcal L}_{kj}\,\varrho_j(t) \right)/\omega
\label{eq:definition_of_rho_tilde}
\ee
as a function of yet unknown operators ${\mathcal L}_{k1}$ to ${\mathcal L}_{kk}$.
According to the initial conditions, the first $k-2$ time derivatives of  $\tilde\varrho_{k+1}({\mathcal L}_{k1},\ldots,{\mathcal L}_{kk},t)$ must vanish at the initial time $t=0$, {\it i.e.}
\be
\frac{\partial }{\partial\hat\varrho_{ij}} \partial^v_t \tilde\varrho_{k+1}({\mathcal L}_{k1},\ldots,{\mathcal L}_{kk},t)|_{t=0} = 0
\hspace*{1cm} \forall i,j
\ee
is required for all $v=0\,\ldots,k-2$.
After the operators ${\mathcal L}_{k1}$ to ${\mathcal L}_{kk}$ have been determined such that they satisfy all these conditions, the auxiliary operator
\be
\varrho_{k+1}(t) \equiv \left(\dot\varrho_k(t) - \sum_{j=1}^{k}{\mathcal L}_{kj}\,\varrho_j(t)\right)/\omega
\label{eq:construction_of_heom_level_k_plus_one}
\ee
vanishes in the first $k$ time-steps as it has been required above.

As this procedure is iteratively re-applied, the number of levels of the HEOM increases until \eqref{eq:construction_of_heom_level_k_plus_one} yields $\varrho_{n+1}(t)=0$.
This indicates that one has obtained a HEOM with $n$ levels, for which the initial condition $\varrho_j(0) = \mathbb{O}$ for $j>1$ gives rise to the targeted system dynamics $\varrho_1(t) = \Lambda_1(t)\,\varrho_1(0)$.

\subsubsection{Example: the derivation of the HEOM \eqref{eq:hierarchyExample2da}/\eqref{eq:hierarchyExample2db}}
\label{sec:appendix_two_level_reviving_coherences}
The method that has been introduced above is now applied in order to motivate how the HEOM in \eqref{eq:hierarchyExample2da}/\eqref{eq:hierarchyExample2db} has been obtained.
To this end, we aim for a HEOM that induces the system dynamics
\be
\label{eq:heom_2d_example_definition}
\varrho_1(t) = \Lambda_1\,{\hat\varrho} = \frac{1}{2}
\left[
\begin{array}{cc}
{\hat\varrho}^{(0)}+{\hat\varrho}^{(z)} & f(t)\left({\hat\varrho}^{(x)} -i {\hat\varrho}^{(y)}\right) \\
f(t)\left({\hat\varrho}^{(x)} + i{\hat\varrho}^{(y)}\right) & {\hat\varrho}^{(0)}-{\hat\varrho}^{(z)}
\end{array}
\right]
\ee
with $f(t)=e^{-\gamma t}\cos(\omega t)$ being the function that characterises the reviving coherences.
The parameters ${\hat\varrho}^{(i)} = \textrm{tr}({\hat\varrho}\,\sigma_i)$ with $i=0,x,y,z$ and $\sigma_0 = \mathbb{I}$ determine the initial state $\hat\varrho$ and need to satisfy ${\hat\varrho}^{(0)} = 1$ and $[\hat\varrho^{(x)}]^2 + [\hat\varrho^{(y)}]^2 + [\hat\varrho^{(z)}]^2 \le 1$ in order for $\hat\varrho$ to have a trace equal to one and to be positive semi-definite.

To find a HEOM that induces the targeted dynamics, we parametrise the yet unknown operator ${\mathcal L}_{11}$ by
\be
{\mathcal L}_{11}\,\varrho = \sum_{i,j} \chi^{\mathcal{L}_{11}}_{ij}\,\sigma_{i}\,\varrho\,\sigma_{j}^\dagger\ ,
\label{eq:general_form_of_L_00}
\ee
where the summation is carried out over $i,j=0,x,y,z$.
In accordance with \eqref{eq:definitition_of_tilde_rho_1}, one defines $\tilde\varrho_2({\mathcal L}_{11},t) \equiv (\dot\varrho_1(t) - {\mathcal L}_{11}\,\varrho_1(t))/\omega$ such that \eqref{eq:determination_of_l_00} reads
\be
\frac{\partial}{\partial\hat\varrho^{(k)}} \tilde\varrho_2({\mathcal L}_{11},0) = 0\hspace*{1cm} \hspace*{5px} \textrm{with} \hspace*{5px}
k=0,x,y,z
\label{eq:determination_of_l_00_example}
\ee
and is solved with respect to $\chi^{\mathcal{L}_{11}}_{ij}$.
This yields $\chi^{\mathcal{L}_{11}}= \textrm{diag}(-\gamma,0,0,\gamma)/2$ or, equivalently, ${\mathcal L}_{11}\,\varrho = \gamma\,{\mathcal D}_z\,\varrho/2$ with the dephasing Lindbladian ${\mathcal D}_z\varrho \equiv \sigma_z\varrho\sigma_z - \varrho$.
\Eref{eq:construction_of_heom_level_k_plus_one} determines the first auxiliary operator $\varrho_2(t)$ as
\be
\varrho_k(t) = 
\left[
\begin{array}{cc}
 0 & g_k(t) \left({\hat\varrho}^{(x)} -i {\hat\varrho}^{(y)}\right)\\
 g_k(t)\left({\hat\varrho}^{(x)} +i {\hat\varrho}^{(y)}\right) & 0 \\
\end{array}
\right]
\label{eq:form_of_auxiliary_operators}
\ee
with $k=2$ and the scalar function $g_2(t) = - e^{-\gamma t} \sin (\omega t)/2$.

We can now re-applying the same procedure to the operator $\varrho_2(t)$, in order to determine ${\mathcal L}_{21}$ and ${\mathcal L}_{22}$.
First, these operators are parametrised in terms of yet unknown scalars $\chi^{{\mathcal L}_{21}}_{ij}$ and $\chi^{{\mathcal L}_{22}}_{ij}$ as in \eqref{eq:general_form_of_L_00}.
After defining
$\tilde{\varrho}_3({\mathcal L}_{21},{\mathcal L}_{22},t) \equiv (\dot\varrho_2(t) - \mathcal{L}_{21}\,\varrho_1(t) - {\mathcal L}_{22}\,\varrho_2(t))/\omega$ as in \eqref{eq:definition_of_rho_tilde},
we solve the conditions
\be
\frac{\partial}{\partial \hat\varrho^{(k)}} \tilde\varrho_3({\mathcal L}_{21},{\mathcal L}_{22},0) = 0
\hspace*{15px}\textrm{ and }\hspace*{15px}
\frac{\partial}{\partial \hat\varrho^{(k)}} \left[\partial_t\tilde\varrho_3({\mathcal L}_{21},{\mathcal L}_{22},t)\big|_{t=0}\right] = 0\ ,
\label{eq:determination_of_l_1k}
\ee
where $k=0,x,y,z$, with respect to parameters $\chi^{{\mathcal L}_{21}}_{ij}$ and $\chi^{{\mathcal L}_{22}}_{ij}$.
The conditions in \eqref{eq:determination_of_l_1k} determine only $24$ out of the $32$ degrees of freedom in $\chi_{ij}^{\mathcal{L}_{21}}$ and $\chi_{ij}^{\mathcal{L}_{22}}$ and the remaining degrees of freedom can be chosen such that the forms of ${\mathcal L}_{21}$ and ${\mathcal L}_{22}$ become as simple as possible, what eventually yields ${\mathcal L}_{21}\,\varrho = \omega\,{\mathcal D}_z\,\varrho/2$ and ${\mathcal L}_{22}\,\varrho = \gamma\,\sigma_z\,\varrho\,\sigma_z$.
The second auxiliary operator
$\varrho_3(t) \equiv (\dot\varrho_2(t) - \mathcal{L}_{21}\,\varrho_1(t) - {\mathcal L}_{22}\,\varrho_2(t))/\omega$ vanishes constantly such that the HEOM has only two level.

We have now derived a HEOM that induces the system dynamics defined in \eqref{eq:heom_2d_example_definition}.
However, we want to generalise this HEOM by permitting an arbitrary pre-factor for ${\mathcal L}_{21}$ such that the latter reads ${\mathcal L}_{21}\,\varrho = \alpha\omega\,{\mathcal D}_z\,\varrho$ with a free parameter $\alpha$.
Furthermore, we permit the damping constant $\gamma$ to take different values $\gamma_1$ and $\gamma_2$ on the two different levels of the HEOM.
For $\alpha = 1/2$ and $\gamma_1 = \gamma_2 = \gamma$ one obtains the original HEOM but the additional parameters also permit to cover different processes.
The final HEOM that has now been obtained is stated in \eqref{eq:hierarchyExample2da}/\eqref{eq:hierarchyExample2db}.

\subsubsection{Example: extension of the HEOM \eqref{eq:hierarchyExample2da}/\eqref{eq:hierarchyExample2db} by means of \eqref{eq:stateHierarchyExample3d}}
\label{sec:extension_of_lindblad}
We can further generalise the dynamics that has been defined in \eqref{eq:heom_2d_example_definition} by modifying the function $f(t)$ to $f(t) = e^{-\gamma t}(1+2\alpha[\cos(\omega t)-1])$.
In the case of $\alpha = 1/2$ one recovers the same dynamics that has been targeted for in \ref{sec:appendix_two_level_reviving_coherences}.
However, different values of $\alpha$ permit a broader range of dynamical processes and call for a new HEOM.

In order to obtain the latter, we follow the proposed procedure in the very same way that has been carried out in \ref{sec:appendix_two_level_reviving_coherences} and find that the first operator in the HEOM reads ${\mathcal L}_{11}\,\varrho = \gamma\,{\mathcal D}_z\,\varrho/2$.
The auxiliary operator $\varrho_2(t)$ is then determined as in \eqref{eq:construction_of_heom_level_k_plus_one} with $k=1$ and is given through \eqref{eq:form_of_auxiliary_operators} with $k=2$ and $g_2(t) = -e^{-\gamma  t}\alpha \sin (\omega t)$.
As in \eqref{eq:definition_of_rho_tilde} we then define $\tilde{\varrho}_{3}({\mathcal L}_{21},{\mathcal L}_{22},t)$  all conditions on which (compare \eqref{eq:determination_of_l_1k}) are satisfied by the operators (the choice is not unique)  ${\mathcal L}_{21}\,\varrho = \alpha\omega\,{\mathcal D}_z\,\varrho $ and ${\mathcal L}_{22}\,\varrho = \gamma\,\sigma_z\,\varrho\,\sigma_z$.
In contrast to the previous example, the operator $\varrho_3(t)$ does in the present case not vanish constantly but is given through \eqref{eq:form_of_auxiliary_operators} with $k=3$ and $g_3(t) = e^{-\gamma t}\alpha (1 - 2 \alpha) (1 - \cos ( \omega  t))$.
For that reason, an additional level in the HEOM is required, which is characterised by the yet unknown operators ${\mathcal L}_{3k}\,\varrho = \sum_{ij}\chi^{\mathcal{L}_{3k}}_{ij}\,\sigma_i\,\varrho\,\varrho_j^\dagger$ with $k=1,2,3$.
By imposing the conditions
\be
\fl
\frac{\partial}{\partial \hat\varrho^{(k)}} \tilde\varrho_4(0) = 0
\textrm{ , }\hspace*{15px}
\frac{\partial}{\partial \hat\varrho^{(k)}} \left[\partial_t\tilde\varrho_4(t)\big|_{t=0}\right] = 0\ ,
\hspace*{15px}\textrm{ and }\hspace*{15px}
\frac{\partial}{\partial \hat\varrho^{(k)} } \left[\partial^2_t \tilde\varrho_4(t)\big|_{t=0}\right] = 0
\label{eq:determination_of_l_3k}
\ee
with $k=0,x,y,z$ on $\tilde{\varrho}_4({\mathcal L}_{31},{\mathcal L}_{32},{\mathcal L}_{33},t)$, which has been obtained from \eqref{eq:definition_of_rho_tilde} with $k=3$, one can eventually determine the operators ${\mathcal L}_{31}\,\varrho = \mathbb{O}$, ${\mathcal L}_{32}\,\varrho = \omega(1-2\alpha)\,\sigma_z\,\varrho\,\sigma_z$, and ${\mathcal L}_{33}\,\varrho =\gamma\,\sigma_z\,\varrho\,\sigma_z$.
Other choices for the operators are possible as \eqref{eq:determination_of_l_3k} does not determine them uniquely, but the present choice has been found most convenient.

Eventually, the HEOM shall again be generalised by replacing the operator ${\mathcal L}_{32}$ with ${\mathcal L}_{32} = \beta\omega\, \sigma_z\,\varrho\,\sigma_z$.
The first two levels of final HEOM are then identical to \eqref{eq:hierarchyExample2da} and \eqref{eq:hierarchyExample2db}.
The latter is, however, extended by an additional summand $\omega \varrho_3(t)$, which evolves according to \eqref{eq:stateHierarchyExample3d}.

\subsubsection{Example: the Jaynes-Cummings model in \eqref{eq:jaynes_cummings_hierarchy_a} - \eqref{eq:jaynes_cummings_hierarchy_c}}
\label{sec:jaynes_cummings}
In contrast to the previous two examples, which were not related to any particular physical system, we will now derive a HEOM that describes the dynamics of the resonant Jaynes-Cummings model \cite{Laine2010}.
To this end, we consider a two-level system inside a leaky cavity whose coupling to the cavity field is determined by the coupling constant $\gamma$.
The field is characterised by a Lorentzian spectral density that is peaked at the resonance frequency of the two-level system and whose spectral width is denoted by $\zeta$.

The reduced system dynamics has been obtained analytically and reads \cite{Laine2010}
\be
\varrho_1(t) = 
\left[
\begin{array}{cc}
|f(t)|^2 \hat\varrho^{(11)} &  f(t) \hat\varrho^{(12)} \\
f^*(t) [\hat\varrho^{(12)}]^* &   1-|f(t)|^2\hat\varrho^{(11)}
\end{array}
\right]\ ,
\ee
where the parameters $\hat\varrho^{(11)} \in \mathbb{R}$ and $\hat\varrho^{(12)} \in \mathbb{C}$ determine the initial system state and 
\be
f(t) = e^{-\frac{\lambda t}{2}}\left[\cosh\left( \frac{\alpha\,t}{2} \right) + \frac{\lambda}{\alpha}\sinh\left( \frac{\alpha\, t}{2}  \right)  \right]
\ee
with $\alpha = \sqrt{\zeta^2-2\gamma\zeta}$ characterises the system dynamics.
Knowing the analytical solution $\varrho_1(t)$, one can apply the procedure that has been described above in order to find a HEOM for the resonant Jaynes-Cummings model.
The derivation is carried out in analogy to \ref{sec:extension_of_lindblad} and eventually yields the HEOM \eqref{eq:jaynes_cummings_hierarchy_a}-\eqref{eq:jaynes_cummings_hierarchy_c}.

\subsection{Extending a Lindblad equation}
\label{sec:extending_of_lindblad_equation}
The HEOMs that have been tailored in \ref{sec:targeted_solution} have been constructed under the premise of a previously defined solution.
One can, however, also take a different phenomenological approach to the construction of HEOMs that is focused on the physical problem rather then on the mathematical form of the solution.
The idea of this approach is to consider a Lindblad equation, which typically permits a convenient phenomenological interpretation, and to extend this Markovian equation to a HEOM that could also permit non-Markovian processes.

As an example, we consider a two-level system that is coupled to a thermal reservoir with finite temperature and whose dynamics is described by the Lindblad equation
$\partial_t\varrho_1(t) = ({\mathcal B}_+ + {\mathcal B}_-) \varrho_1(t)$ with
$
{\mathcal B}_\pm\varrho_1 = \gamma_\pm \left(\sigma_\pm\,\varrho_1\,\sigma_\mp- \left\lbrace \sigma_\mp\sigma_\pm,\varrho_1 \right\rbrace /2 \right)
$
and rates $\gamma_\pm$.
Depending on the initial state, the dynamics that is induced by this Markovian equation of motion is characterised by exponential gain or loss process.

This Markovian equation can now be extended such that it becomes a HEOM by adding $\omega\,\varrho_2(t)$ to the right-hand side of the equation of motion, where $1/\omega$ is a unit of time and $\varrho_2(t)$ denotes the (dimensionless) auxiliary operator.
The equation of motion for $\varrho_2(t)$ shall be of the form $\partial_t \varrho_2(t) = {\mathcal L}_{21}\,\varrho_1(t) + {\mathcal L}_{22}\,\varrho_2(t)$ such that the two operators ${\mathcal L}_{21}$ and ${\mathcal L}_{22}$ determine to what extent the system dynamics will eventually differ from what would have been induced by a time-dependent Lindblad equation.
How one can uncover an intuitive relation between the choice of ${\mathcal L}_{21}$, ${\mathcal L}_{22}$ and system dynamics is a challenging and in full generality yet unanswered question.
A good phenomenological understanding can, however, be obtained for the case of ${\mathcal L}_{21}$ being proportional to ${\mathcal L}_{11}$, i.e.\ for ${\mathcal L}_{21} = (\xi/\gamma_+)({\mathcal B}_+ + {\mathcal B}_-)$ with a free scalar parameter $\xi$. 
This choice of ${\mathcal L}_{21}$ causes the system state to affect the time-derivative $\partial_t\varrho_2(t)$ of the auxiliary operator in the same way (up to a constant factor) as the time-derivative $\partial_t\varrho_1(t)$ of the system state itself.
If there was no further coupling between $\varrho_1(t)$ and $\varrho_2(t)$, then the time-evolution of $\varrho_1(t)$ would be completely encoded into the time-evolution of $\varrho_2(t)$.
However, due to the summand $\omega\,\varrho_2(t)$ in the equation of motion for $\varrho_1(t)$, the auxiliary operator $\varrho_2(t)$ couples back to the system state and affects the latter not at the time $t$ but only in the next time-step $t + \delta t$.
This contributes a kind of ``inertia'' to the system dynamics and gives rise to non-Markovian oscillation of quantum coherence.

Eventually, we choose ${\mathcal L}_{22}$ such those coherences between $\sigma_x$-eigenstates and those between $\sigma_y$-eigenstates in the auxiliary operator are damped with the rate $(\gamma_+ + \gamma_-)/2$.
This gives rise to ${\mathcal L}_{22}\,\varrho = (\gamma_+ + \gamma_-) ( {\mathcal D}_x + {\mathcal D}_y )/2$ and the HEOM that has been stated in \eqref{eq:bath_heom_a}-\eqref{eq:bath_heom_b}.

\end{appendix}

\section*{References}
\providecommand{\newblock}{}

\end{document}